\documentclass[fleqn,usenatbib,useAMS]{mnras}

\usepackage{graphicx}	
\usepackage{amsmath}	
\usepackage{amssymb}	
\usepackage{multicol}        
\usepackage{bm}		
\usepackage{pdflscape}	
\usepackage{journals}

\newcommand{\Msun}{M\ensuremath{_\odot}\,}

\usepackage{etoolbox}
\makeatletter
\newcount\c@additionalboxlevel
\newcount\c@maxboxlevel
\patchcmd\@combinedblfloats{\box\@outputbox}{%
    \stepcounter{additionalboxlevel}%
    \box\@outputbox
}{}{\errmessage{\noexpand\@combinedblfloats could not be patched}}

\AtBeginShipout{%
    \ifnum\value{additionalboxlevel}>\value{maxboxlevel}%
    \typeout{Warning: maxboxlevel might be too small, increase to %
        \the\value{additionalboxlevel}%
    }%
    \fi
    \@whilenum\value{additionalboxlevel}<\value{maxboxlevel}\do{%
        \typeout{* Additional boxing of page `\thepage'}%
        \setbox\AtBeginShipoutBox=\hbox{\copy\AtBeginShipoutBox}%
        \stepcounter{additionalboxlevel}%
    }%
    \setcounter{additionalboxlevel}{0}%
}
\makeatother

\def\HI{H~{\sc i}\, }
\def\HII{H~{\sc ii}\, }

\def\kms{$\textrm{km~s$^{-1}$}$}
\def\nb{\textsc{nbursts}}

\usepackage[T1]{fontenc}
\usepackage{ae,aecompl}

\usepackage{newtxtext,newtxmath}
\usepackage{epstopdf}

\title[]{Spectral observations of the systems with the disturbed spiral arms: Arp~42, Arp~82 and Arp~58}
\author[Zasov et al.]{
Anatoly V. Zasov,$^{1,2}$\thanks{E-mail:zasov@sai.msu.ru}
Anna S. Saburova,$^{1}$
Oleg V. Egorov,$^{1,3}$
Sergei N. Dodonov $^3$
\\
$^1$ Sternberg Astronomical Institute, Moscow M.V. Lomonosov State University, Universitetskij pr., 13,  Moscow, 119234, Russia\\
$^2$ Faculty of Physics, Moscow M.V. Lomonosov State University, Leninskie gory 1,  Moscow, 119991, Russia \\
$^3$ Special Astrophysical Observatory, Russian Academy of Sciences, Nizhniy Arkhyz, Karachai-Cherkessian Republic, 357147, Russia\\
}

\begin{document}
\large

\label{firstpage}
\pagerange{\pageref{firstpage}--\pageref{lastpage}} \pubyear{2019}
\maketitle

\begin{abstract}
We study three Arp' systems of peculiar galaxies: Arp~42 (NGC~5829), Arp~82 (NGC~2535/36) and Arp~58 (UGC~4457), using the long-slit spectral observations carried out at the 6m telescope BTA of Special Astrophysical Observatory. Arp~82 and Arp~58 are the M51-type systems.  In the third system -- Arp~42, there are two  extremely luminous kpc-sized clumps observed at the ends of the bifurcated spiral arm, however the source of perturbations  remains unknown. From the emission line measurements we analyzed the distribution of gas line-of-sight (LOS) velocity, velocity dispersion and gas-phase metallicity along the slits estimated  by different methods. A special attention is paid to the young regions of  star formation at the peripheries of the galaxies and in tidal debris, their connection with gas kinematics and abundance. All three systems show the signs of velocity disturbances and  the presence of regions of locally enhanced gas velocity dispersion, as well as the regions of diffuse ionized gas. A distribution of oxygen reveals shallow radial abundance gradients, typical for interacting systems.   A faint spiral-like branches of tidal bridges which are observed in NGC~2535 and UGC~4457 may represent the remnants of pre-existing "old" mode of spiral waves.
\end{abstract}

\section{Introduction}

Tidal interactions of spatially close gas-rich galaxies affect mostly their gas component due to the collisional nature of gas and low dispersion of gas velocities. Some fraction of  gas of interacting galaxies loses angular momentum and moves closer to the center, usually stimulating a burst of star formation, while some gas and stars may be  thrown out of galactic discs, often forming tidal streams stretching far away from parent galaxies. At the stage of strong interaction, direct collisions of gas flows can occur leading to shock-induced star formation (see e.g. \citealt{Maji2017} and references therein).  Formation of stars is often observed both in a far periphery of disc and in tidal debris of galaxies including a formation of gravitationally bound tidal dwarf galaxies, which takes place under favourable conditions \citep[see the surveys by][]{Weilbacher2002, Ducetal2013}. 

 Mechanisms which can promote a formation of the extended star-forming regions at the periphery of galaxies and in tidal debris in low gas density conditions, as well as their subsequent evolution, is the  actively debated problem \citep[see e.g.][]{Wetzstein2007, Bournaud2011}. Obviously, concrete scenarios depend on many factors, and they are different for different interacting systems. In addition to the gravitation instability of gas in tidal debris, which does not always take place, a key role in triggering star formation may play such events as a supersonic turbulence, ram pressure and interaction with the external gas \citep{Renaud2014, Maji2017, Smith2013, Kapferer2008, Tonnesen2012}. So, as it was noted earlier by \citet{Boquien2009}, collisional debris may be considered as laboratories to study star formation.  
 
 It is essential that the galaxies we consider and their tidal structures are in the very unstable phase of evolution when stellar and gas spatial distributions and their velocities may experience a significant change over a characteristic dynamic time of about $10^8$ yr which complicates the observed picture. Although strong interaction affects all regions of a galaxy, it manifests itself in different ways at different distances from the  centre: due to gravity torques the gas looses angular momentum in the inner galaxy and gains it at larger radii \citep{Bournaud2011}.  As a result there is a local increase of gas density and activation  of star formation in the circumnuclear  region of a galaxy \citep{Noguchi1988, Hernquist1989}, and a decrease in gas density in other regions, which in turn may cause a flattening of radial gradient of gas abundance \citep{Kewley2006, Kewley2010}.  Gas and stars ejected from the disc may either leave the galaxy  or fall backward \citep[see, e.g.][]{Elmegreen1993}, they may form a tidal tail or tidal dwarf galaxies \citep{Duc2000, Ducetal2013}. Measuring gas velocities and  radial gradients of chemical composition in interacting systems allows us to understand better the star formation conditions and the ways of the evolution of galaxies, which have experienced or experienced strong tidal disturbances in the past.

In this paper we continue to describe the results of spectral observations at the 6m telescope BTA of Special Astrophysical Observatory of disturbed galaxies with local regions of star formation in the periphery of discs or in the intergalactic space which we started in \citet{zasovetal2015,zasovetal2016,Zasovetal2017,Zasovetal2018}.
Here we focus on the  three Arp systems Arp 42, Arp 82, and Arp 58. The main components of  these systems  are the late-type star-forming spiral galaxies with peculiar spiral structures (a presence of tidal arms or arm bifurcation), evidencing their perturbed state. The galaxies  are characterized by active star formation, where the numerous local sites (clumps) of emission gas and young stars are noticeable not only in their central parts, but also at the far disc periphery and in tidal debris.  Composite SDSS {\it g,r,i}- band images of the galaxies are shown in Fig. \ref{map}.

The structure of this paper is as follows.
In the next section we give the short information on the considered systems; the observation and data reduction procedures are presented in Section 3; 
it is followed by Section 4 describing the data analysis techniques we use; 
the results of observations are given in Section 5, and their discussion is presented in Section 6. Section 7 summarizes our conclusions.

\section{A short information on the chosen galaxies}
\subsection{Arp 42} 
Arp 42 is the poorly studied system, consisting of Sc-type galaxy NGC~5829 and the irregular-looking IC4526 observed at the projected distance of 2.3\arcmin, or about three times of its optical radius $R_{25}$ = 0.7\arcmin.  R-band surface photometry of this system was carried out by \citet{Reshetnikov1993}. Curiously, the satellite galaxy is most probably a background object in spite of its peculiar asymmetric shape and the extended  faint outer arm looking like a  tidal tail (see Fig \ref{map}). Indeed, the systemic velocities  $V_{GSR} $ of NGC~5829 and IC~4526 are 5759 \kms\, and 13732 \kms\, respectively  \citep[HYPERLEDA\footnote{http://leda.univ-lyon1.fr/} ,][]{Makarov2014},  evidencing that these galaxies are not interacting ones. Since there are no other objects of comparable brightness in the vicinity, hereafter we will consider Arp 42 as a single object NGC~5829, assuming its redshift-based distance to be 78 Mpc. It is a gas-rich late-type galaxy. Its total mass of neutral hydrogen is about $2\times10^{10}$\Msun (Hyperleda; see also \citealt{Casasola2004}). Mass of stellar population may be found from r-band luminosity using the model ratio $M/L_r \approx 2.0$ \citep{Bell2003} for the color index $(g-r)=0.54$ corrected for Galactic extinction estimated from SDSS images. The resulting value is $M\approx  5\times  10^{10}$ \Msun.

NGC~5829 is a Grand Design-type  galaxy, its two arms look non-identical. The southern spiral arm  contains bright clumps  of \HII. Remarkably, this arm contains two or three straight sections (`rows'). Such features often observed in galaxies with the ordered spiral arms may be considered as a sign of strong large-scale shock waves linked with spiral arms \citep[see][ and the references therein]{Cherninetal2001, Butenkoetal2017}. In contrast, the opposite (northern) spiral arm looks  rather smoothed and badly defined, although the extended emission regions are observed along the whole arm. Its outer fraction is bifurcated, and both  split components contain a similarly looking kpc-sized unusually bright compact emission regions (we'll call them BER1,2, see Fig. \ref{map}), embedded into the area of the enhanced brightness of about 6 kpc length, slightly  stretched along their parent  spiral branches.

\subsection{Arp 82}
Arp 82=VV~009 is the system of M51-type. It consists of a luminous spiral galaxy  with the well defined two-armed spiral structure (NGC~2535), one arm of which connects or overlaps with the close satellite (NGC~2536). In addition, NGC 2535 possesses a long tidal tail as the continuation of the north spiral arm at the opposite side from the satellite.  The systemic velocity of the main galaxy is close to 4000 \kms, so we adopt the distance 54 Mpc for this system. 

Both galaxies ~NGC~2535/36  in the system Arp~82 are  embedded in the large \HI envelope evidently associated with the main galaxy. The extended gaseous disc is too large to be the result of the observed interaction, and most probably it existed before the galaxies became close  \citep{Kaufman1997}. For the adopted distance a total mass of \HI in the galaxy pair  is $2.3\times 10^{10}M_\odot$ (in agreement with single dish observations) where only a  small portion of \HI ($5\times10^8M_\odot$) belongs to the satellite \citep{Kaufman1997}, however the latter overlaps with the gaseous tidal bridge extending from NGC~2535, which makes the  \HI mass estimation model-dependent. 

A resolved kinematics of Arp~82 was studied by several authors. 2D velocity map was obtained by \citet{Kaufman1997} in \HI line and by \citet{Amram1989} in $H\alpha$ line, using a scanning Fabry--Perot interferometer. \citet{Klimanov}  also obtained long-slit velocity measurements for two slit position angles (59\degr ~and 163\degr)  for NGC~2535 and its satellite. All authors indicate  a strong non-circular gas motions caused by interaction and the distortion of the velocity field due to the bending of the outer disc of NGC~2535, although the rotation curve of the central disc of this galaxy (within R$\sim$ 20--25~arcsec) is fairly symmetric. The velocity field distortion and the uncertainty in evaluation of inclination angles of both galaxies does not allow to obtain a reliable dynamic estimate of their masses. \citet{Kaufman1997} also pointed out the enhanced turbulent velocities of \HI allover the system (about 30 \kms\, along the LOS).

Numerical modeling of interacting galaxies with prograde rotation, whose masses are comparable \citep{Howard1993} or differ by several times \citep{Holincheck2016, Kaufman1997} reproduces successfully the main observed  features of Arp~82. Velocity field of gas  also well agrees with the  Klaric' model (see discussion in \citealt{Kaufman1997}). The model of interacting galaxies presented by  \citet{Hancock2007} indicates that the galaxies have experienced two close encounters, and the most recent encounter have caused the observed burst of star formation in both galaxies.

\subsection{Arp 58}
Arp 58 = UGC~4457 = VV~413 also belongs to the M51-type systems. Its main component is the Sc-type two-arm mildly inclined galaxy. The satellite is observed at the end of the  SE-spiral arm at a distance of about one optical diameter $D_{25}\sim$1~arcmin. The second spiral arm at the opposite side from the satellite has a faint extension which forms a thin  curved tidal tail. Systemic velocity of the main galaxy is close to 11000 \kms, and the distance adopted in this paper is 147 Mpc. 

UGC~4457 is very luminous galaxy: its corrected B- magnitude is $B_{tc}$ =  14.24 (HYPERLEDA database), which  corresponds to luminosity 6.6$\times 10^{10}L\odot$ for the adopted distance.  The satellite is a few tens of times fainter than the main galaxy.  This small galaxy is observed at a stage of active star formation.    Its inner part has an elliptical shape and demonstrates  strong emission lines in its spectrum, however the dim outer regions of the companion are of  irregular shape being distorted by interaction.

\section{Observations and data reduction }\label{obs}

\begin{figure}
 \vspace{-2.6cm}
\hspace{-1.0cm}
\includegraphics[width=1.15\linewidth]{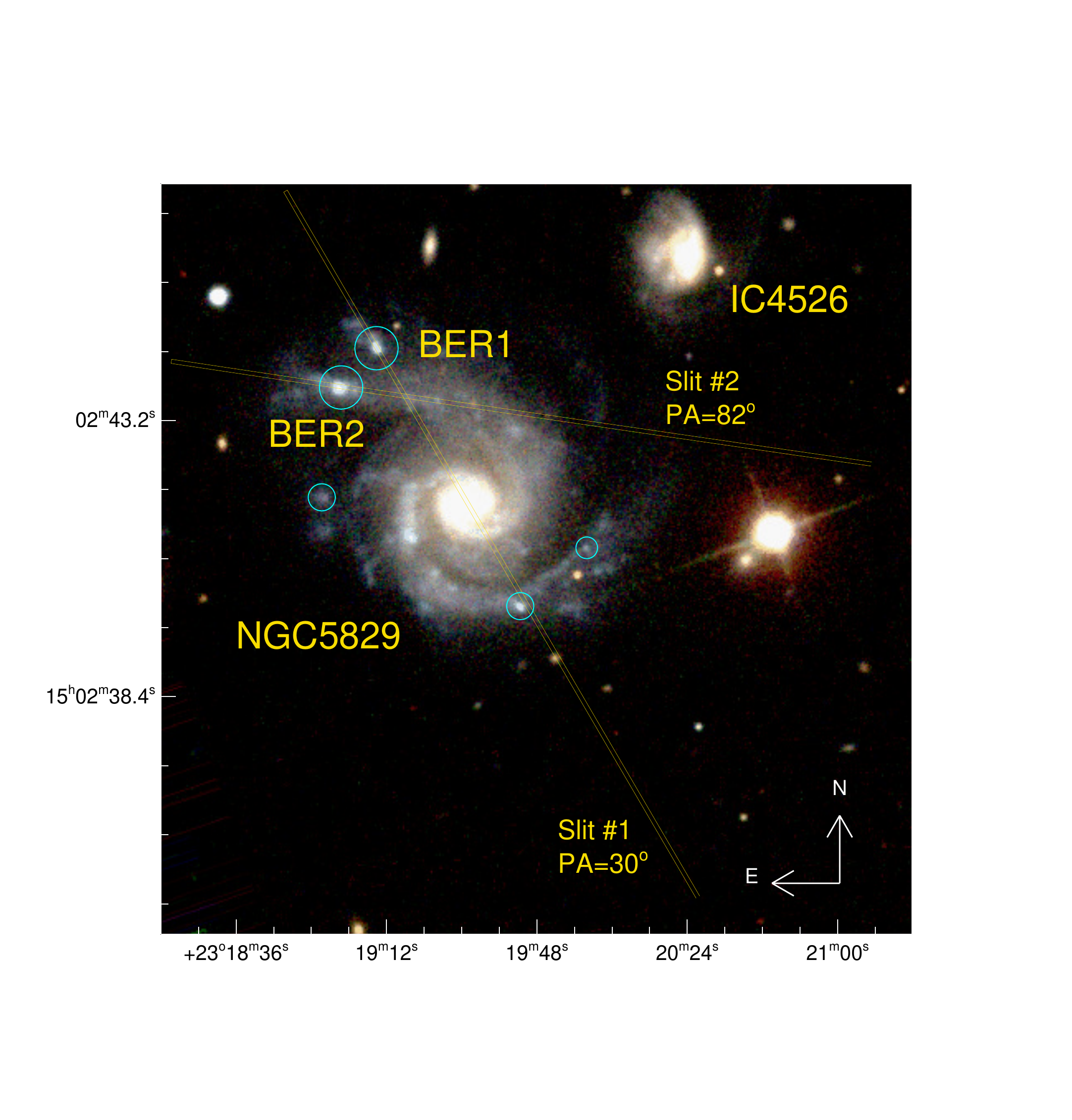}

 \vspace{-2.6cm}
\hspace{-1.0cm}
\includegraphics[width=1.15\linewidth]{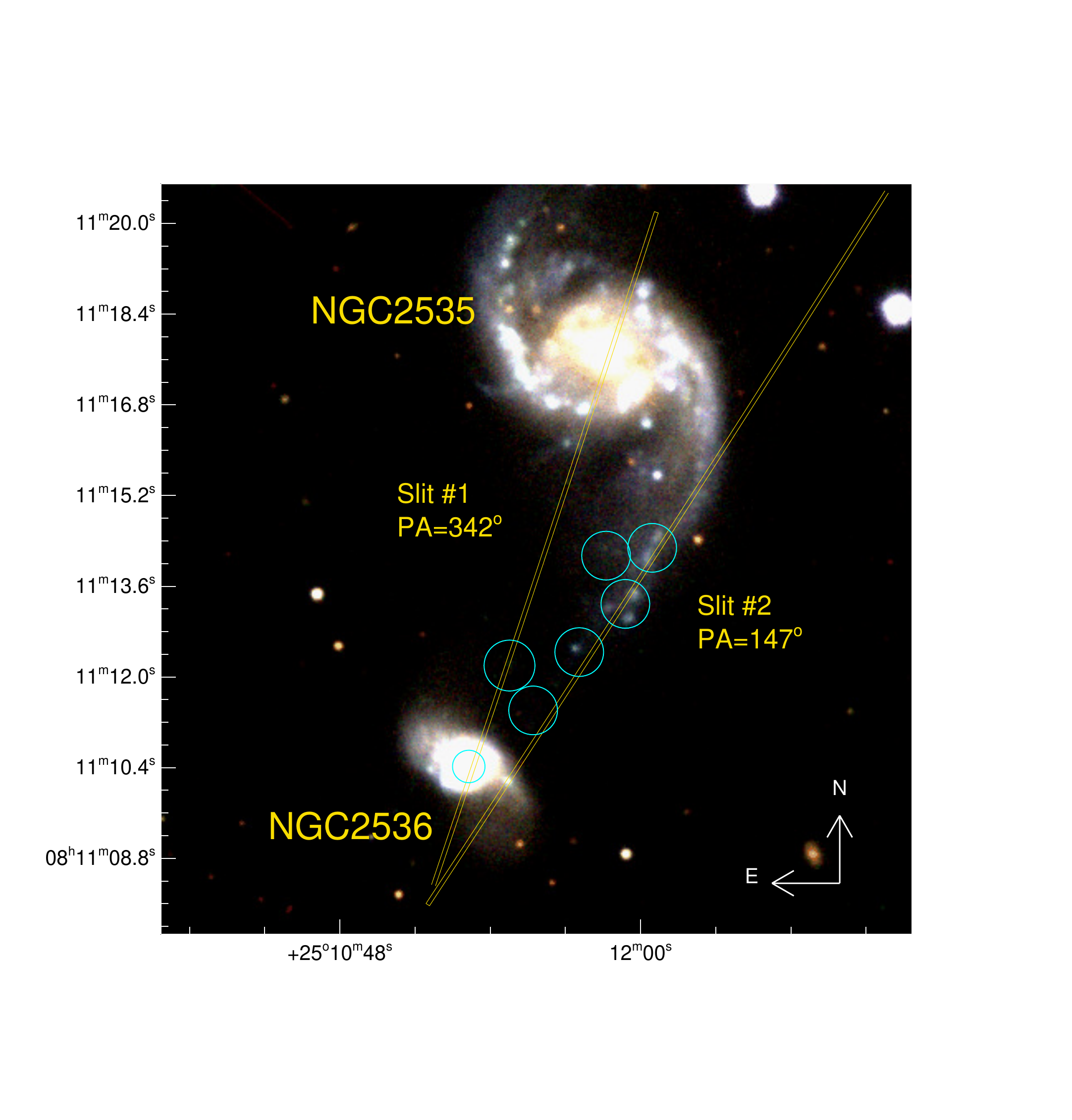}

 \vspace{-2.6cm}
\hspace{-1.0cm}
\includegraphics[width=1.15\linewidth]{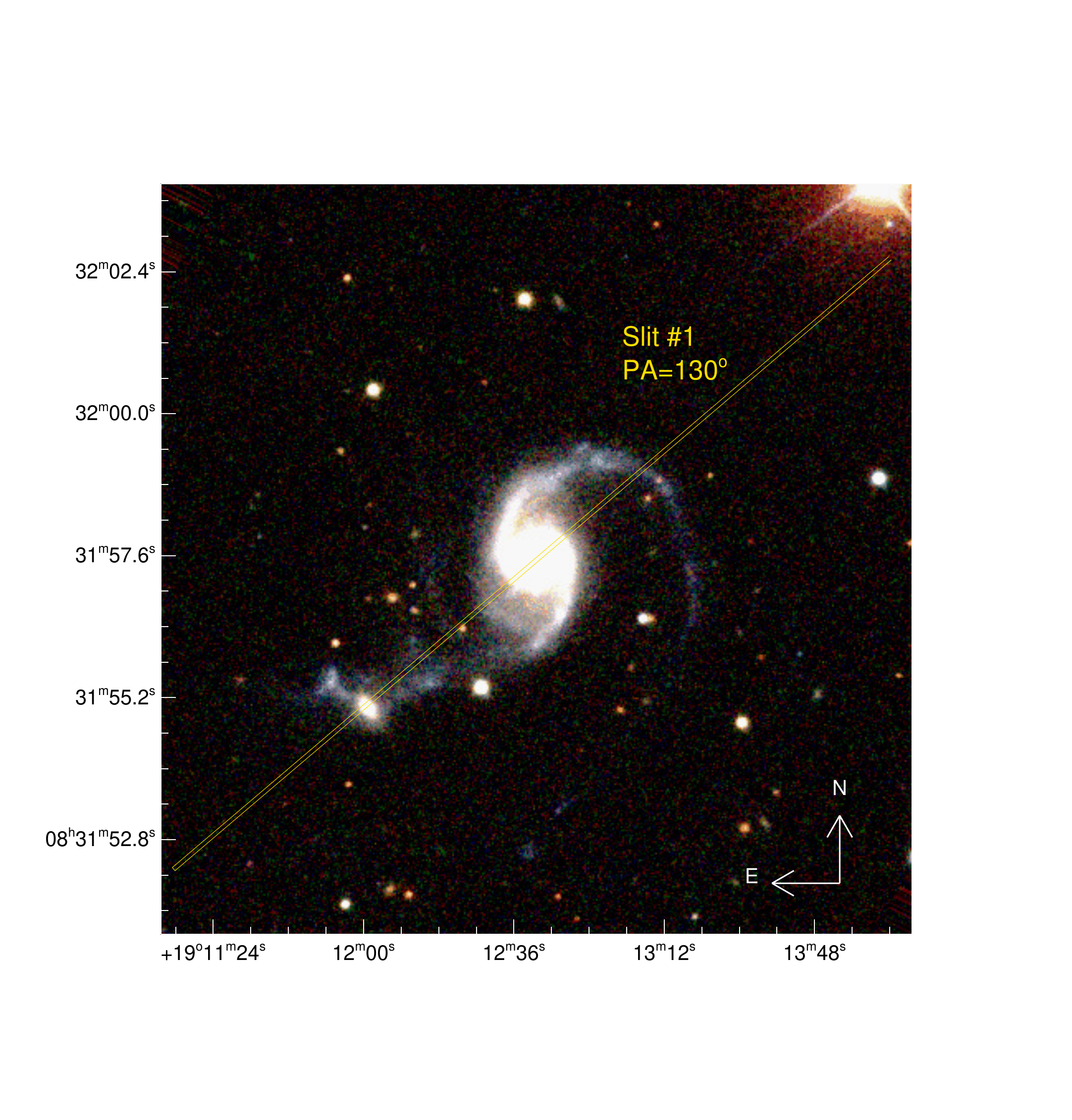}
\vspace{-1.2cm}
\caption{The positions of the slit used for spectroscopic observations overplotted on the composite SDSS {\it g,r,i}- band images of  Arp~42, Arp~82, Arp~58 (from top to bottom). The circles on the images of Arp~42 and Arp~82 mark the positions of sites of star formation that were considered in the two-coloured diagrams (see below).}
\label{map}
\end{figure}

{The long-slit spectral observations of Arp~42, Arp~82 and Arp~58 were performed with the Russian 6-m telescope with SCORPIO-2 spectrograph \citep{AfanasievMoiseev2011}. We give the log of observations in Table \ref{log}. For every galaxy it contains a position angle (PA) of the slit; date of observations; the total exposure time and seeing. We utilized the grism VPHG1200@540, which covers the spectral range 3600--7070 \AA\, and has a dispersion of 0.87 \AA\ pixel$^{-1}$. The spectral resolution is $\approx 5.2$ \AA\, estimated as FWHM of night-sky emission lines}. The scale along the slit was 0.36~arcsec pixel$^{-1}$,  and the slit width was  1 arcsec. Slit positions are superimposed on the optical images of the galaxies in Fig.~\ref{map}. Every slit was chosen to cross the bright  star-forming clumps.

A procedure of data processing was described step-by-step in our earlier papers \citep{zasovetal2015, zasovetal2016,Zasovetal2017, Zasovetal2018}. 
To process the data we used the \textsc{idl}- based pipeline. Briefly, the data reduction consisted of the following stages: a bias subtraction and truncation of overscan regions, flat-field correction, the wavelength calibration based on the spectrum of He-Ne-Ar  lamp, cosmic ray hit removal, summation of individual exposures, the night sky subtraction and flux calibration using the spectrophotometric stellar standards BD33d2642, BD28d4211, BD25d4655. 

\section{Data analysis}\label{sec:analysis}
 In our spectra analysis we took into account the parameters of  instrumental profile of the spectrograph resulted from the fitting  of the twilight sky spectrum observed in the same  observation runs. We convolved these parameters with the high-resolution PEGASE.HR~\citep{LeBorgneetal2004} simple stellar population models (SSP) and fitted the reduced spectra of galaxies.  We performed it using the \nb{} full spectral fitting technique  \citep{Chilingarian2007a, Chilingarian2007b}, which  allows to fit the spectrum in a pixel space.  In this method the parameters of the stellar populations are derived by nonlinear minimization
of the quadratic difference chi-square between the observed and model spectra. We utilized the following parameters of SSP: age T~(Gyr) and metallicity [$Z/H$]~(dex) of stellar population. The line-of-sight velocity distribution (LOSVD) of stars was parameterized by Gauss-Hermite series (see \citealt{vanderMarel1993}).

The emission spectra were obtained by subtraction of model stellar spectra from the observed ones. After that we fitted the Gaussian profiles
to emission lines to estimate the velocity and velocity
dispersion of ionized gas and the fluxes in the emission
lines.

To increase the signal-to-noise ratio (S/N) of the spectra we used the adaptive binning  in the fitting. The S/N threshold levels we   specified for Arp~42, Arp~82 and Arp~58 are  15, 30 and 20 respectively. We used this binning to evaluate kinematics.  To estimate the emission line fluxes and metallicities we utilized binning of 3 pixels  for every spectrum.  The pixels with S/N $< 3$ in emission lines were excluded from further pixel-by-pixel analysis. 

\begin{figure}
\hspace{-0.4cm}
\includegraphics[width=1\linewidth]{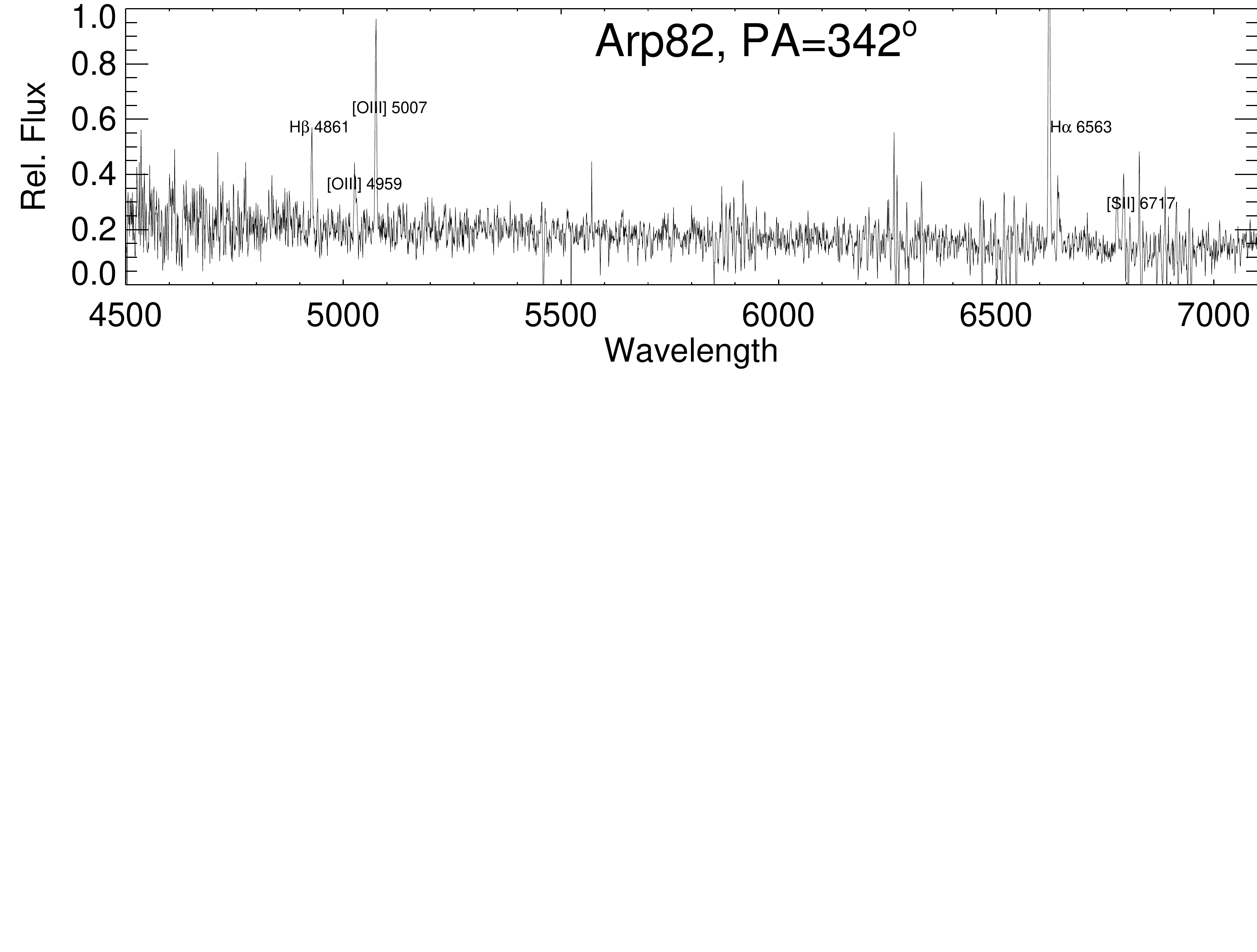}
\vspace{-3.5cm}
\hspace{-0.4cm}
\includegraphics[width=1\linewidth]{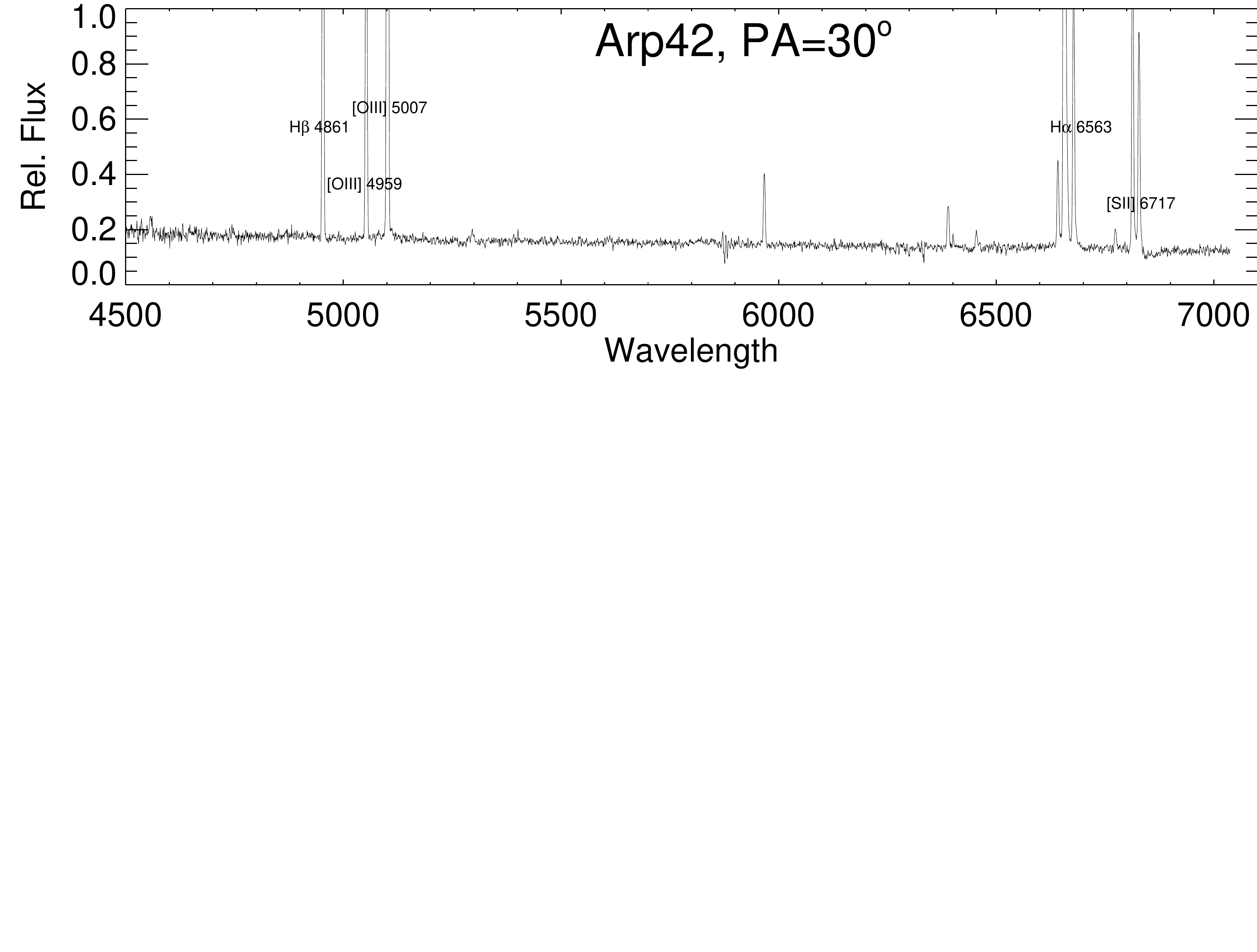}
\vspace{-4.0cm}
\caption{The examples of stacked spectra of the low (top panel) and high (bottom panel) S/N bins used for the metallicity estimate at $R\approx -80$ arcsec  for Arp~82 and BER1 for Arp~42 respectively.}
\label{spectrum}
\end{figure}

In addition, we also analyzed the fluxes and intensities from a stacked spectra of several adjoining pixels corresponding to single star-forming clumps or to extended regions of fainter emission. A choice of the pixels interval to be stacked was made by eye in order to increase S/N to the values not less than 10 for bright emission lines; a typical size of such bins is close to the size of bright clumps crossed by slits, or about $10-15$~arcsec for faint extended regions. The resulting estimates for these bins (shown by squares in  Figures below) are close to those obtained from pixel-by-pixel analysis with much lower S/N and thus prove the reliability of the latter.   For illustrative purposes we demonstrate in Fig.\ref{spectrum} the examples of spectra of the stacked bins with low and high S/N that we used in our estimates of metallicity (see below). The first one is the spectrum of the region with the coordinate $R\approx -80$ arcsec in Arp~82 ($PA=342$\degr), and the second one is for BER1 in Arp~42  ($PA=30$\degr).

To calculate a gas-phase metallicity, we use several methods of estimation of (O/H) abundances from the relative intensities of strong 
  emission lines. They are based either on the empirical calibrations by comparison with the oxygen abundance values obtained by the $T_e$-method  (S-method by \citealt{Pilyugin16}, O3N2 method by \citealt{Marino2013}), or on the photoionization models of \HII regions (\textsc{izi} by \citealt{Blanc2015} made using \citealt{Levesque2010} photoionization models).  In the most cases the shape of radial profiles of $12+\log\mathrm{(O/H)}$ obtained by different methods looks rather similar, although  there is a well known discrepancy between the theoretical and empirical estimates: theoretical models usually lead to higher values of abundances  \citep[see for example][]{Kewley2008}. 
  
  Note that the regions showing the signs of non-photoionization mechanism of the ionized gas excitation  were not used for estimates of the oxygen abundance.  The conclusion whether a region shows non-photoionization mechanism of excitation was made based on the position on the so-called   BPT diagnostic diagrams proposed by \citet{BPT} and extended by \citet{BPT_S} for the [O~\textsc{iii}]/H$\beta$ vs [S~\textsc{ii}]/H$\alpha$ case. These diagrams are widely used to separate the different mechanisms of emission lines excitation. Additionally, all regions with equivalent width of H$\alpha$ line EW(H$\alpha$) < 3 (according to \citealt{Lacerda2018}) were excluded as probably related to the diffuse ionized gas (DIG). In the mentioned paper the authors claimed that EW(H$\alpha$) > 14 \AA\, is typical for pure star-forming region. However applying  another criterion to separate DIG from \HII regions -- based on $H\alpha$ flux \citep{Zhang2017} -- changes the results negligibly. Moreover, \citet{Kumari2019} showed that estimates of the metallicity made using O3N2 method doesn't suffer from the presence of the DIG. Hence our analysis based on at least this method should be reliable even if the DIG contribution in the remaining data is still significant.

\section{Results of  observations}

\subsection{Arp~42}

For Arp~42 we obtained two spectral cuts for two slit orientations. The slit positions are demonstrated in Fig. \ref{map} (top panel).
In the first cut ($PA_1 = 30$\degr) ~the slit crosses the center of galaxy and BER1.  This slit orientation differs at about 12\degr ~from the optical major axis (MA) $ (PA)_0\approx 18$\degr ~\citep[HYPERLEDA;][]{Nishiura2000}. In the second cut $PA_2 = 82$\degr ~the slit passes along the second  branch of the spiral arm crossing the southern BER2.  

\begin{figure*}
\includegraphics[width=0.5\linewidth]{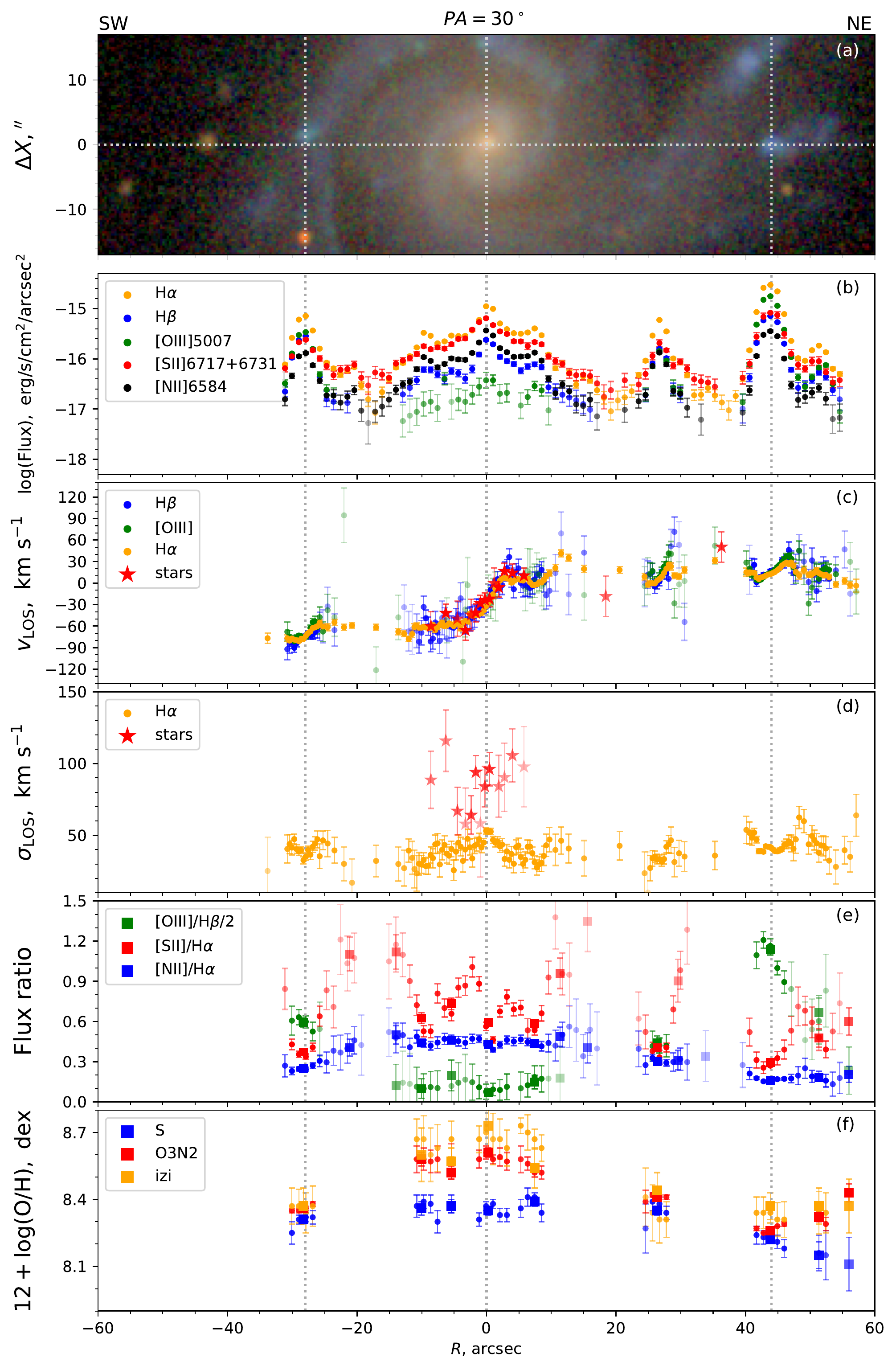}
\includegraphics[width=0.5\linewidth]{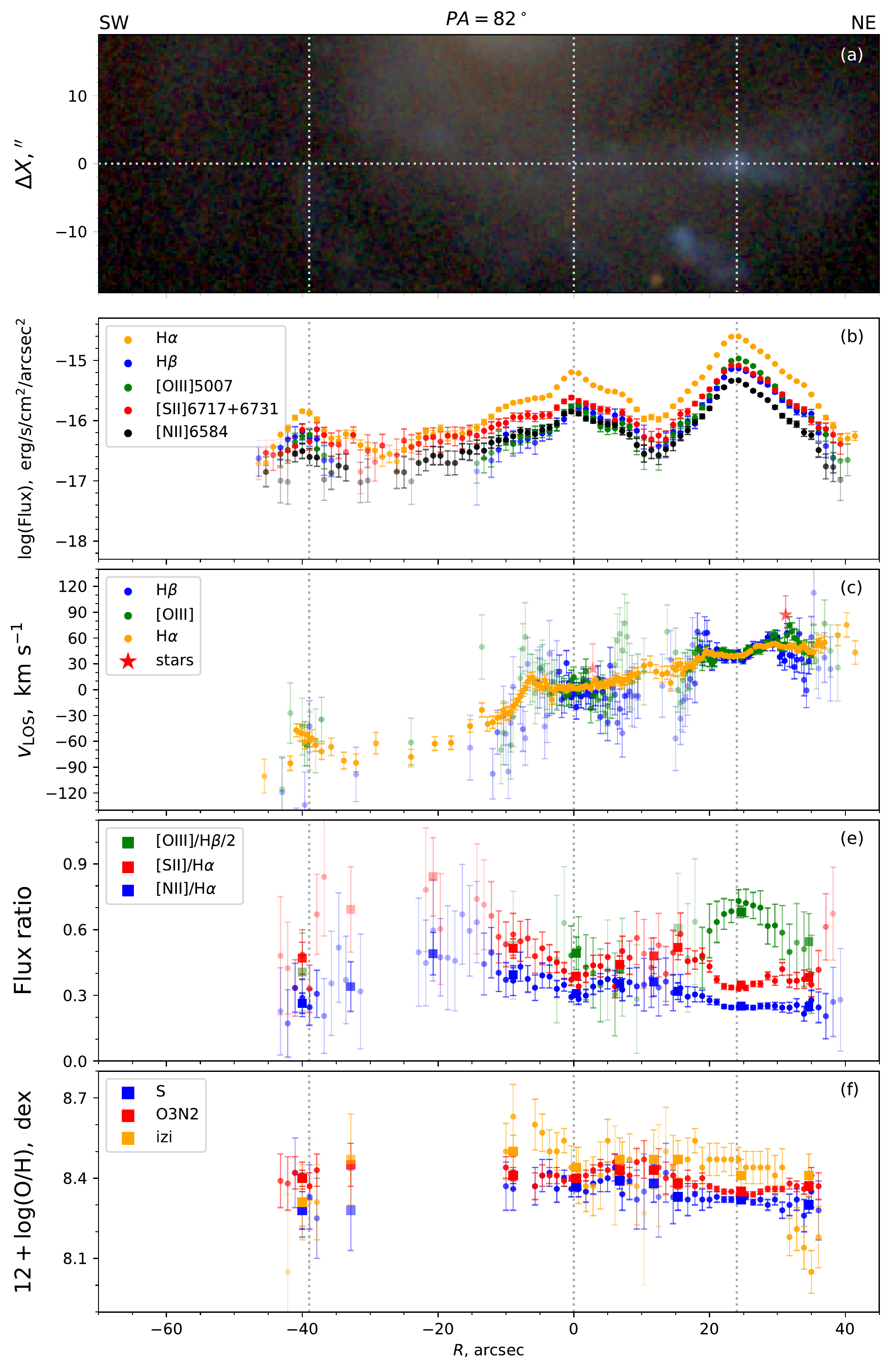}
\caption{The radial variation of measured parameters for $PA_1=30$\degr ~(left panel) and $PA_2=82$\degr ~(right panel) for Arp~42. From top to bottom: the slit positions overlaid on the SDSS \textit{gri}-images (a);  observed emission lines fluxes (b); LOS velocity (c); LOS velocity dispersion (d); flux ratios (e); and oxygen abundance (f). Circles correspond to the different emission lines. Stars demonstrate the stellar kinematical data. Square symbols show the values obtained after stacking of several adjoining pixels corresponding to single star-forming clumps or to extended regions of fainter emission (see Section~\ref{sec:analysis}). For clarity we mark the estimates having large uncertainties relative to other points at a given plot by pale-colored symbols.}
\label{arp42_results}
\end{figure*}

The results of data processing are illustrated in Fig. \ref{arp42_results}   for $PA_1=30$\degr ~(left panel) and $PA_2=82$\degr ~(right panel). There we  show the slit position overlaid on the composite SDSS {\it g,r,i}- band images (a) and  from top to bottom the radial variation of: (b)  observed emission lines fluxes; (c) LOS velocities; (d)  LOS velocity dispersion for $PA=30$\degr\; (e)  flux ratios and (f) oxygen abundance. Circles correspond to the different emission lines. For convenience we mark the estimates having large uncertainties relative to other points at a given plot by pale-colored symbols.  Stars in Fig.~\ref{arp42_results} demonstrate the stellar kinematical data obtained from the absorption spectra. Square symbols show the values obtained after stacking of several adjoining pixels corresponding to single star-forming clumps or to extended regions of fainter emission (see Section~\ref{sec:analysis}). For reference we mark the position of some morphological features (\HII clumps, galaxies centres etc.) by vertical dotted lines. Note that for $PA=82$\degr~ we managed to estimate only the upper limit of LOS velocity dispersion in all emission lines (50 km~s$^{-1}$), so we do not reproduce it in the Figure.

LOS velocity profile along $PA_1= 30$\degr~ in general agrees with that  obtained earlier by \citet{Nishiura2000} along major axis, although it extends to larger radial distances R and is more detailed. Its shape in general reflects the shape of the rotation curve: it grows in the inner few kpc and then flattens at larger R.  Local non-circular motions with the amplitude up to several tens of \kms\, are observed  where the slit crosses bright emission regions.

\begin{table}
\caption{Log of observations}\label{log}
\begin{center}
\begin{tabular}{ccccc}
\hline\hline
&Slit PA & Date & Exposure time& Seeing \\
 &  ($^o$)  &  &     (s) &        (arcsec) \\
\hline
Arp~42&30&07.04.2016&1800&1.5 \\
Arp~42&82&24.04.2015&4500&3.3 \\
\hline
Arp~82&342&12.12.2015&7200&4.0 \\
Arp~82&147&13.12.2015&6300&4.0 \\
\hline
Arp~58&130&12.12.2015&5400&1.4 \\
\hline
\end{tabular}
\end{center}
\end{table}

\begin{figure}
\includegraphics[width=\linewidth]{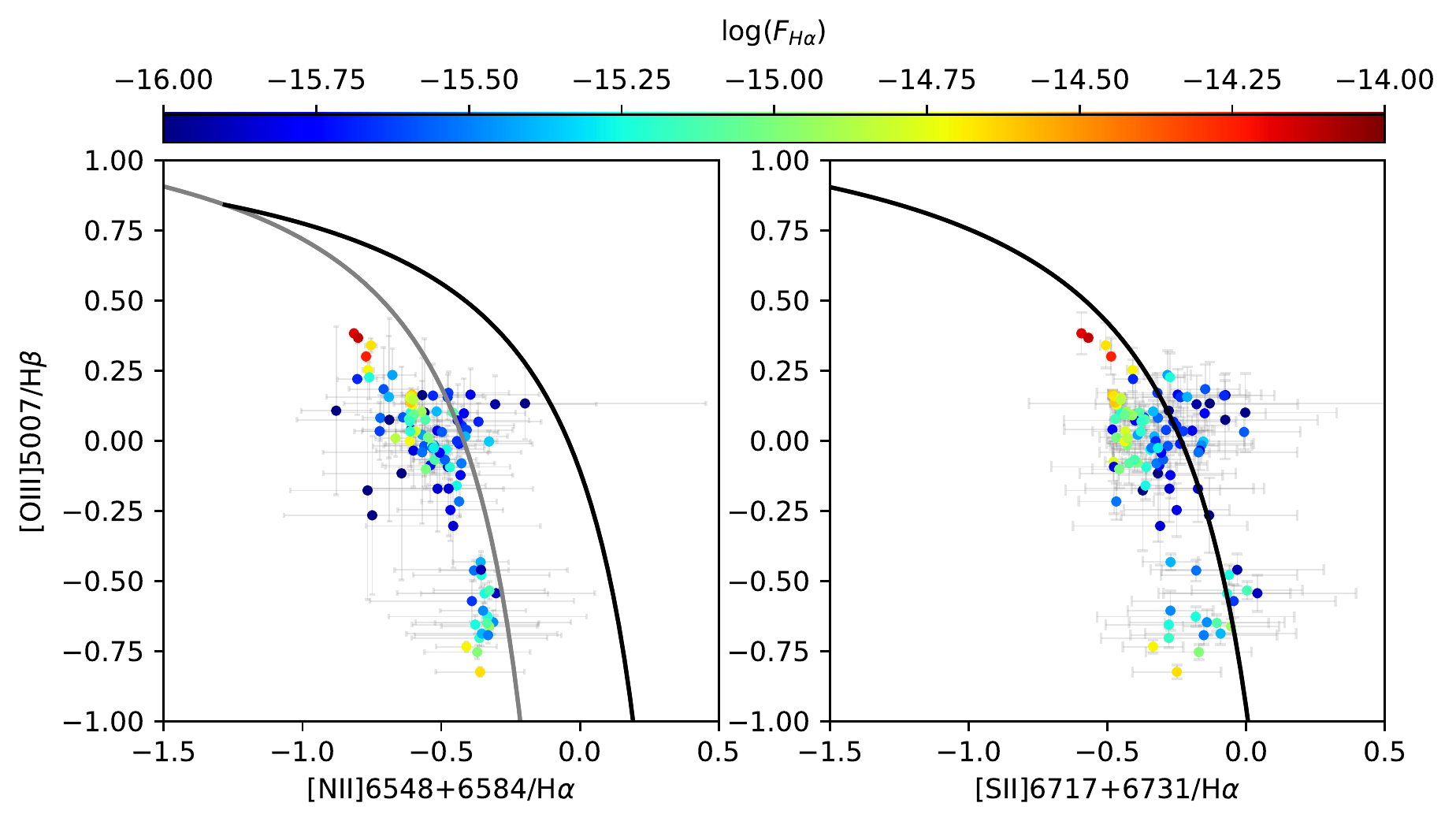}
\caption{{BPT-diagrams plotted for individual bins along the slit  for Arp~42. The `maximum starburst line' \citep{Kewley2001} separating photoionised \HII-regions and all other types of gas excitation is shown in black; grey line from \citet{Kauffmann03} separates the regions with composite mechanism of excitation. Colour denotes the $H_\alpha$-flux.}}
\label{arp42_bpt}
\end{figure}
\begin{figure*}
\includegraphics[width=0.5\linewidth]{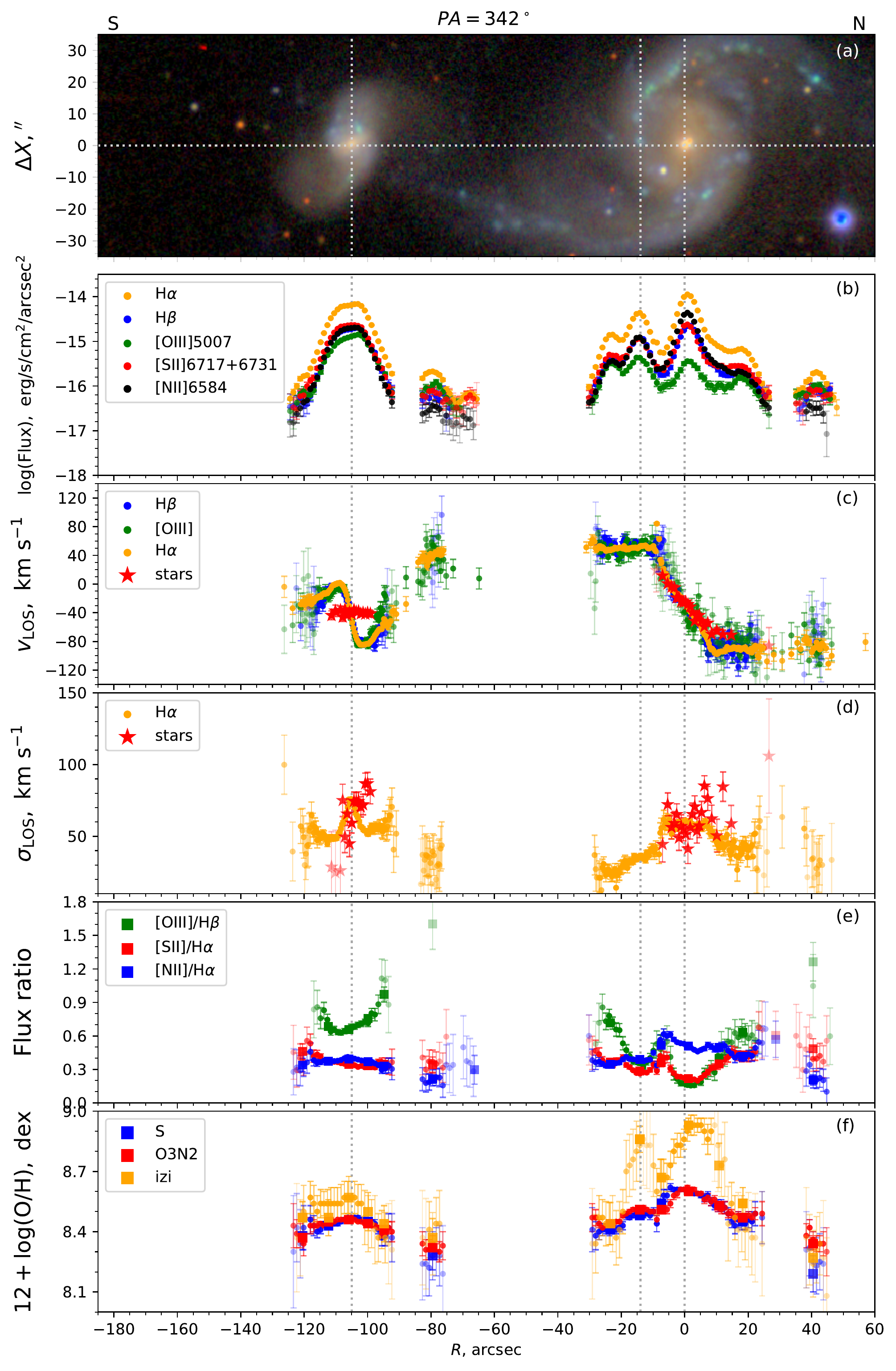}
\includegraphics[width=0.5\linewidth]{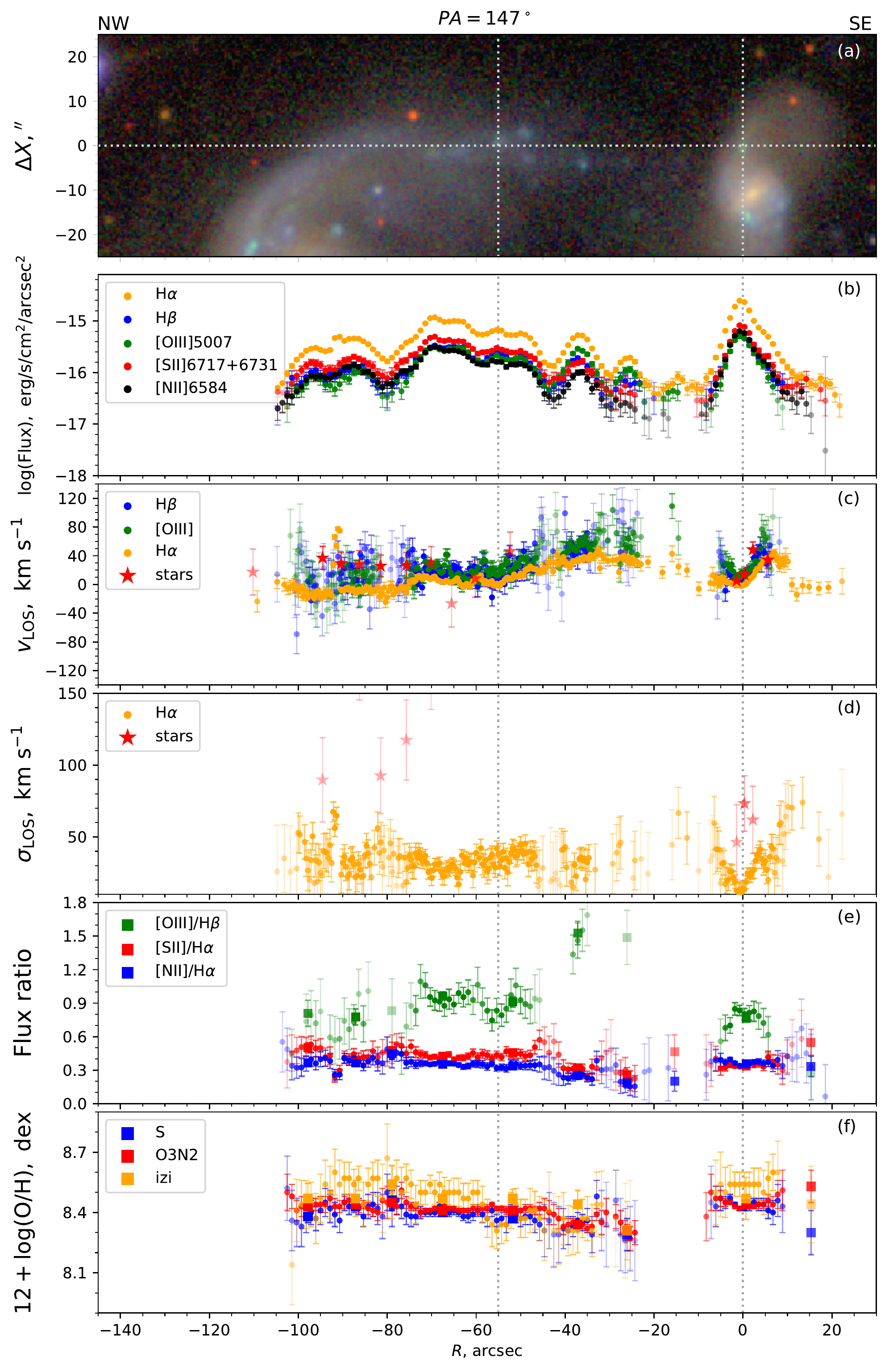}
\caption{Same as in Fig.~\ref{arp42_results}, but for Arp~82 $PA=342$\degr~(left) and $PA=147$\degr ~(right)}.

\label{arp82_results}
\end{figure*}

\begin{figure}
\includegraphics[width=\linewidth]{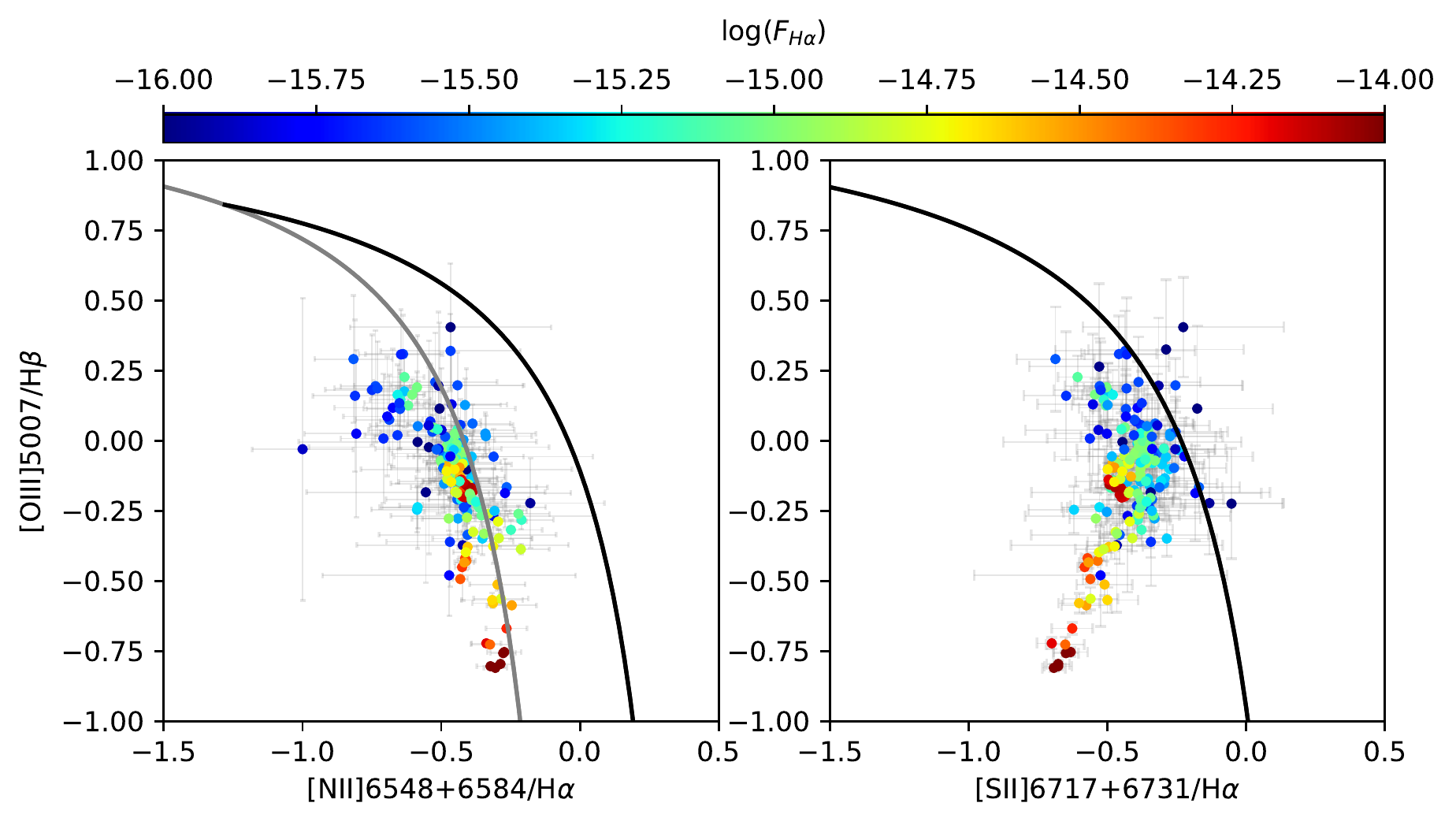}
\caption{Same as in Fig.~\ref{arp42_bpt}, but for Arp~82.}
\label{arp82_bpt}
\end{figure}

\begin{figure}
\includegraphics[width=\linewidth]{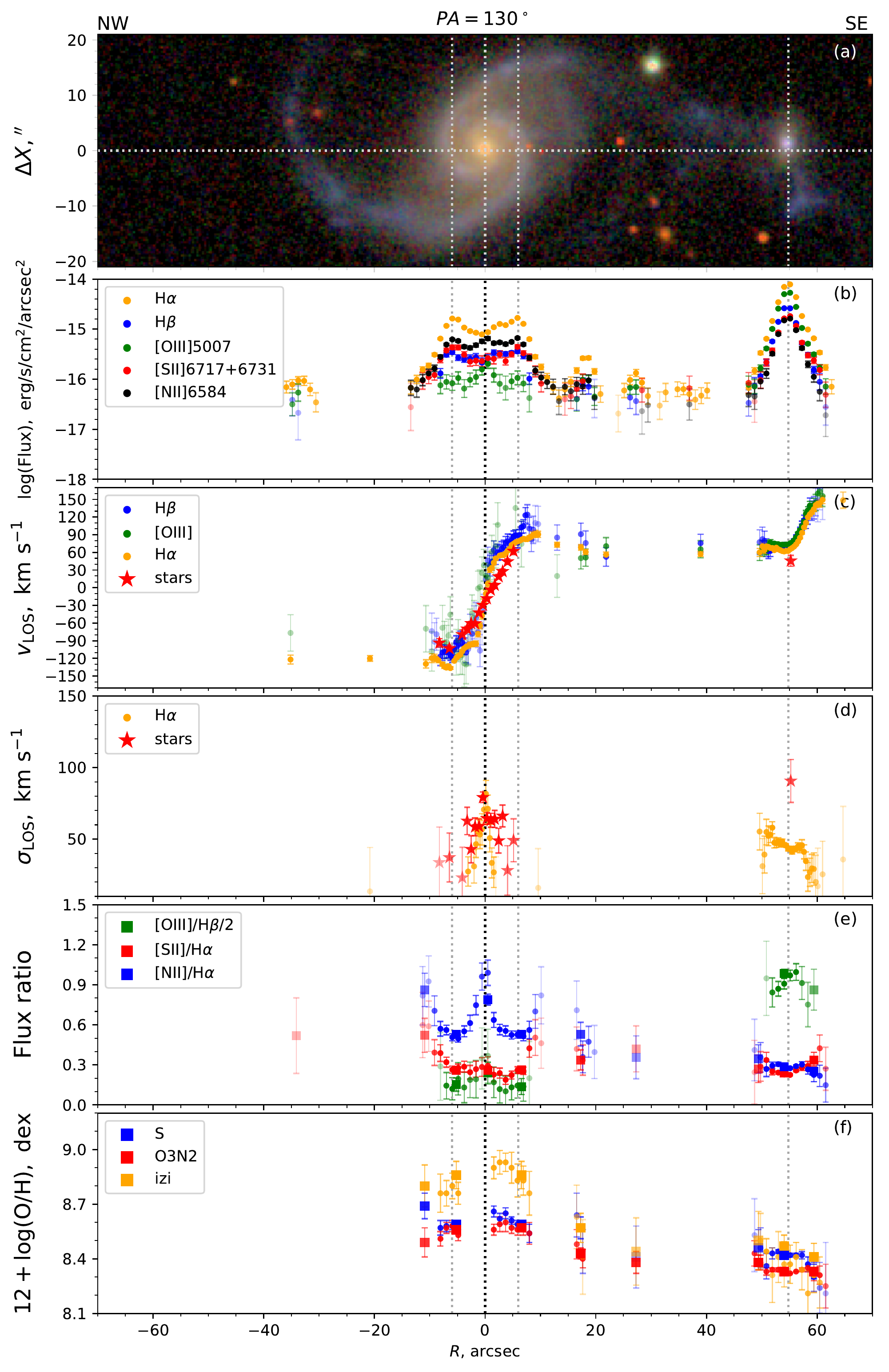}
\caption{Same as in Fig.~\ref{arp42_results}, but for Arp~58.}
\label{arp58_results}
\end{figure}

\begin{figure}
\includegraphics[width=\linewidth]{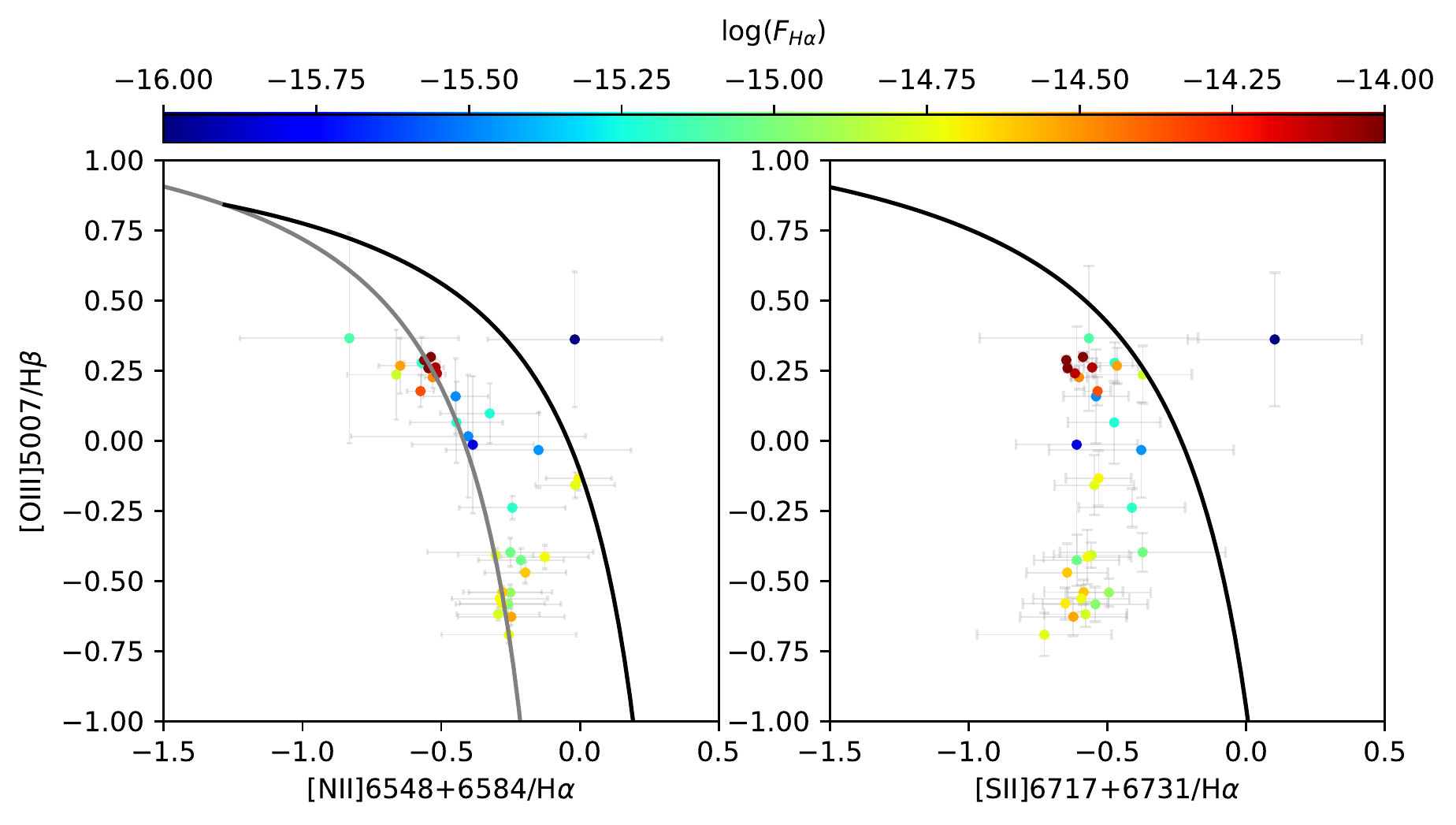}
\caption{Same as in Fig.~\ref{arp42_bpt}, but for Arp~58.}
\label{arp58_bpt}
\end{figure}

Velocity dispersion of emission gas does not exceed 50 \kms, which is normal for galaxies with moderate star formation rate (here and below in this paper we provide the velocity and velocity dispersion estimates only for the regions with S/N>15).     Stellar velocity dispersion $c_*\sim$ 80-100 \kms ~ was measured only  for the central part of the galaxy. This value\	 is quite normal for late-type galaxies of similar luminosities with moderate bulge (e.g. NGC~1084, NGC~4254, see the HYPERLEDA database \citep{Makarov2014}).

In Fig. \ref{arp42_bpt} we demonstrate the position of the emission regions crossed by the slit in the  BPT diagnostic diagram,  where the black `maximum starburst line' \citep{Kewley2001} separates the photoionized H~\textsc{ii}-regions from all other types of gas excitation (AGN, shocks etc.). A composite mechanism of excitation is expected for points between the black line and the gray line proposed by \citet{Kauffmann03} (see left panel). One can see that most of the dots for Arp~42 lie in the pure photoionization region of the diagram. Several points beyond the borderlines for photoionized gas belong to regions of low surface brightness in the strong emission lines, evidencing the significant input of non-photoionizing mechanism of gas excitation.  

Oxygen abundance that we found by different methods (see Fig. \ref{arp42_results}, bottom panels)  shows a wide spread of values along PA=30\degr. The empirical O3N2 and theoretical \textsc{izi} methods give higher values 
for the centre of the galaxy. At the same time at the effective radius of the disc $r_e=10.8$ kpc our metallicity estimate is close to that found in the CALIFA survey \citep{califa2016} for NGC~5829 ($\log\mathrm{(O/H)}=8.43\pm 0.05$).  For the second  cut, which crosses the periphery of NGC~5829, the metallicity remains roughly constant $12+\log\mathrm{(O/H)}\approx 8.4$. Oxygen abundance does not  show any change when the slit passes through the BER1 and  BER2.

\subsection{Arp~82}

We fulfilled two spectral cuts of Arp~82 (see Fig.\ref{map}, central panel). Slit PA =342\degr ~passes through the centres of galaxies. This orientation is close to (PA)$_0\approx 153$\degr ~of major axis (MA) for the outer elliptical optical isophotes of the galaxy as well as for the kinematic MA \citep{Kaufman1997}. The second slit PA=147\degr ~runs along the bridge between the galaxies and cuts the western part of NGC 2536. The slit PA=147\degr ~also crosses the region of splitting of the spiral arm of NGC~2535 about half-way between galaxies, clearly visible in GALEX image (see Section~\ref{sec:disc}).

Profiles of LOS velocities, velocity dispersion, line intensities and oxygen abundance are presented for both positions of the slit in Fig. \ref{arp82_results}.  LOS velocities profile along PA=342\degr ~shows that the maximum velocity of rotation of NGC~2535 corresponding to the flat parts of rotation curve is $V_{rot}\sin i\approx$ 70 \kms, in a good agreement with \citet{Amram1989} data, who observed this system in $H_\alpha$ line.  

Unlike NGC~2535, velocity profile along the diameter of its satellite has a peculiar shape: LOS velocity of gas  grows with radial distance, but at R>5~arcsec (1.4 kpc)  its sign reverses with respect to the centre, so at R>10~arcsec the rotation changes its direction to the opposite one at both sides from the centre.   This result slightly differs from \citet{Amram1989} data, which also reveal the rapid decrease of rotation curve, however the velocity does not cross a zero. The kinematical profile of the satellite obtained by \citet{Klimanov}  agrees in general with our data. Such velocity profile may be explained by strong bending of the gaseous disc, so that at $R\approx$10~arcsec disc inclination passes through zero. However  most probably it reflects non-circular ordered gas motions due to stellar bar, clearly visible in the center of this galaxy. Unlike gas velocities, the stellar LOS velocities remain constant along the radius. To clarify the gas and stellar motions in this region, 2D velocity field is needed.

Velocity profile along the second slit PA=147\degr ~confirms a very low velocity change along the bridge between the galaxies (at about 20-30 \kms). When crossing the spiral arm of NGC~2536, a velocity gradient is observed which may be accounted for galaxy rotation.
Stellar velocity profile slightly disagrees with the gaseous one along the slit: it is 20-40 \kms higher than the latter in the bridge and remains less disturbed  where the slit crosses the blurred spiral arm of the satellite.

Velocity dispersion of gas is relatively low for both spectral cuts: judging from H$\alpha$ data, which give the most precise estimates, it does not exceed 50 \kms\, with the exception of the central region of both galaxies, where small central bars are observed, and probably of the outer low brightness regions of the satellite (PA=147\degr). 
Between the galaxies as well as  in the low-contrast spiral arm-bridge velocity dispersion of emission gas has the lowest values (about 30 \kms), which coincides with the \HI velocity dispersion according to \citet{Kaufman1997}. 

Note that in the main body of NGC~2535 the gas velocity dispersion decreases along the radius parallel with the brightness of emission lines, which suggests that the  kinetic energy of random gas motion in the emission regions is connected with the intensity of local star formation there \citep[see also][for  similar correlation in other galaxies]{Moiseevetal2015}. 

Of special interest are the areas of low brightness of emission lines crossed by slits. 
As it follows from the diagnostic diagrams, (see Fig.\ref{arp82_bpt}),  both faint and bright emission regions,  have a photo-ionization as the main mechanism of  excitation. However there are few exceptions where the input of collisional excitation, enhancing the [S~\textsc{ii}] vs H$_\alpha$ ratios,  may be significant. It is noteworthy that all these cases relate to faint emission regions with low $H_\alpha$-flux values. 

An oxygen abundance of gas was found by several different methods based on the relative intensities of bright emission lines. The results are shown in Fig \ref{arp82_results} (bottom left and right panels). The (O/H) distribution along the line connecting the centers of two galaxies (Fig \ref{arp82_results}, bottom left-hand panel) reveals a significant negative gradient of (O/H) along the slit within the distance $\sim 20$ arcsec  ~from the nucleus of NGC~2535 especially for  the \textsc{izi} method. At the same time the profile (O/H) along PA= 147\degr ~evidences the nearly constant abundance along the tidal bridge: (O/H) decreases at about 0.2 dex within 80 arcsec, or 22 kpc length. In accordance with expectation, gas metallicity in the bridge between galaxies corresponds to metallicity of the outer regions of the main galaxy confirming that it consists of gas expelled from the disc periphery.

\subsection{Arp~58}

For this system we obtained one spectral cut with the slit passing at a distance ~2 arcsec~ (which is close to the seeing) from the center of the satellite (see Fig. \ref{map}, bottom panel). The results of the measurements are illustrated in Fig. \ref{arp58_results}.

The difference of systemic velocities of two galaxies in the system is about 60 \kms, and the highest LOS velocity of the main galaxy overlaps with the lowest velocity of its satellite at the side faced to UGC~4457. Nevertheless the profiles of LOS velocities of both galaxies are different.  The main galaxy, unlike its companion, demonstrates a regular rotation. The measured velocity gradient  of stellar population in the central part of the galaxy is a little lower than for  the emission gas, which may be accounted for higher stellar velocity dispersion in the centre.  In contrast,  a velocity profile of the companion is very asymmetrical. It reveals nearly constant velocity in the NW part of the satellite (towards the main galaxy), which is close to the stellar velocity in the centre of galaxy, and a steep velocity gradient on the opposite side from the centre.  It can indicate the outflow of gas or a strong distortion of gaseous layer, which is not unexpected  if to take into account the very disturbed appearance of this galaxy. 

We managed to estimate velocity dispersion of gas and stars for the central parts of galaxies only. As in the other two Arp' systems it does not exceed 50 \kms for the emission gas. Curiously, stellar velocity dispersion for the main galaxy is lower than for its satellite evidencing the absence of a significant bulge there.

Oxygen abundance distribution along the slit demonstrates a decreasing of (O/H) with the distance from the centre of UGC~4457, however different methods of abundance estimate we used give different absolute values of  (O/H) and its gradient (see the  Discussion below).

Fig.  \ref{arp58_bpt} compares line ratios for different regions of Arp~58. Their positions at the BPT diagrams correspond to the
normal \HII regions formed by young stars or to the composite mechanism of excitation. The only outlier point belongs to the center of satellite.

\section{Discussion}\label{sec:disc}

Below we discuss the main results of photometric and spectral data analysis for three systems in question separately.

\subsection{Arp~42}

 Dynamic mass of the spiral galaxy NGC~5829 (a component of Arp 42) may be estimated only roughly.  Assuming the photometrically defined inclination $i \approx 48$\degr ~given by HYPERLEDA, we obtain the velocity of rotation of the outer disc $ V_{rot} $ of about 110-135 \kms. It is in agreement with HYPERLEDA database, which  gives $V_{rot}= 115 \pm 3$ \kms ~ based on the line width of \HI. Total mass within the optical radius may be taken as $M(R_{25}) = k\times V_{rot}^2R_{25}/G$ where k=1 for spherical mass distribution and~0.6 for the flat disc model \citep{Nordsieck1973}. In the former, most expected,  case $M(R_{25})\approx (6\pm 1)\times10^{10}M_\odot$.  It agrees with the total stellar mass  $M_* \approx 5\times 10^{10}M_\odot$ found from  photometry of {\it g} and {\it r}-band images.

 Oxygen abundance in the inner part of the galaxy is $12+\log\mathrm{(O/H)}$  = 8.4-8.6 for O3N2 and \textsc{izi} methods, and 8.3--8.4 for S method. It is not in conflict with its stellar mass, at least for O3N2 method (see, e.g., the stellar mass-oxygen abundance relationship for O3N2 method in \citealt{Sanchez2017}), and also agrees with the high relative mass of gas (H+He): $\mu=\dfrac{M_{gas}}{M_*+M_{gas}}\approx 0.35$ (see  Fig. 10 in \citealt{Hughesetal2013}).
 
 Oxygen abundance does not reveal a flattening of radial profile $\log$(O/H) at least up to $R_{25}$ (see Fig \ref{arp42_results}), which is often observed in interacting galaxies \citep[see f.e.][]{Rosa2014}. All methods we used demonstrate a smooth decreasing of O/H along the radius with gradient 0.05-0.1 dex at $R_{25}$, see Section \ref{oh}. The directly observable  gradient is lower along the slit PA = 82$\degr $ because this slit does not pass through the galactic centre. 
 
As it was noted in the Introduction, the most spectacular features of gas-rich spiral galaxy NGC~5829 are two extended  emission regions containing two bright clumps BER1,2, in the bifurcated northern spiral arm (see Fig.\ref{map}).  

We estimated colour indices of compact stellar regions BER1,2 from the photometry of SDSS images and corrected them  for Galactic extinction  and internal extinction calculated from the $H_\alpha/ H_\beta$ ratio: $E_{B-V}\approx 0.4$.Their g-magnitudes taken with the surrounding bright emission area of several kpc-size,   are 17.1$^m$ and  17.6$^m$ (not corrected for the extinction), while g-magnitudes of the compact clumps BER1, BER2 within the 4 arcsec aperture are  18.2$^m$ and 17.8$^m$ respectively,  which corresponds to luminosities (4-6)$\times10^8L_\odot$. 
Taking into account their blue \textit{ g-r} colours  0.12, 0.16 respectively (corrected for Galactic extinction  according to \citet{Schlafly2011}), their stellar masses in the model \citep{Bell2003} are (2-3)$\times10^8M_\odot$. It exceeds typical masses of globular clusters, being closer to a  a mass of a dwarf galaxy (for the usually accepted IMF). Accounting for total extinction in g,r-bands has little effect on the determination of the stellar mass, since it increases the estimate of the luminosity and about the same decreases the mass to luminosity ratio.

At any rate, BER1,2 are much more luminous than so-called stellar super-clusters observed in many strongly interacting or merging galaxies rich of gas, such as Antennae where they have properties expected for young massive globular clusters \citep{Whitmore2010}.  The most  luminous of the Antennae' superclusters have absolute magnitude $M_V= -(13-14.6)$, corresponding to luminosity (1-6)$\times10^7L_\odot$  - that is they are much fainter than BERs!  To avoid the conclusion about the enormously big mass of these stellar clumps one can admit a bottom-light IMF of their stellar population.

Note however, that the Antennae system is at a distance of 22 Mpc (against a distance of 78 Mpc for Arp 42), and the mentioned observations \citep{Whitmore2010} were carried out with very high quality (HST).   SDSS resolution does not allow to resolve the images of bright clumps, but they look definitely not as point-like sources. The aperture 4 arcsec which was applied to the clumps corresponds to the diameter about 1.5 kpc. Evidently, both BER1,2 are the combination of several bright young stellar complexes (superclusters), surrounded by  the extended zone of several kpc size of emission gas and active star formation.

\begin{figure}
\hspace{-1.2
cm}
\includegraphics[width=1.3\linewidth]{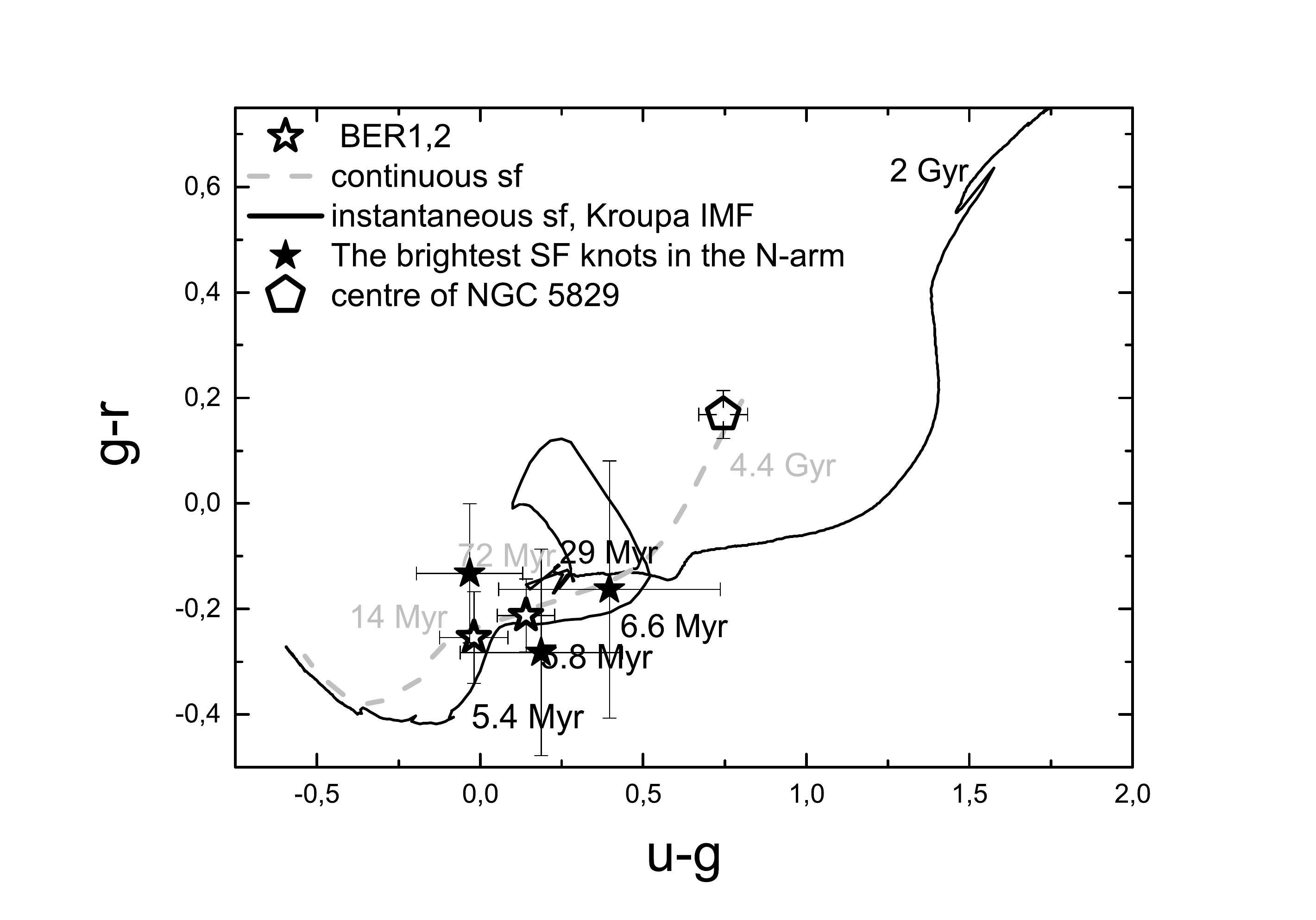}
\caption{The (g-r) vs (u-g) diagram for NGC~5829. Open stars show the positions of BER1 and BER2. Filled stars and hexagon relate to the bright local sites of star formation in the opposite spiral arm and to the central part of NGC~5829.  Black line and gray dashed line  follow  Starburst99 model tracks for instantaneous and continuous star formation respectively. We mark the ages of stellar population corresponding to the closest regions of the tracks by black and gray printing type for instantaneous and continuous star formation models correspondingly. 
}
\label{arp42_col}
\end{figure}

 Extinction-corrected positions of BER1,2   are represented as open stars in the two-colour diagram shown in Fig.\ref{arp42_col}. They  are in satisfactory agreement with  Starburst99 \citep{Leitherer1999} model tracks for both continuous and instantaneous star formation  \citep[for Kroupa IMF][]{Kroupa2001}  and  confirm their very  young stellar age: $t=5-14$~Myr. In turn, the central part of the galaxy (open hexagon),  better fits with the model for continuous star formation  and evidently consists of   a mixture of stars of different ages, so its luminosity-weighted age of stellar population is about 4 billions yr. For the comparison we also overplot the position of three  most luminous clumps of star formation in the opposite spiral arm of NGC~5829 (filled stars), their position is shown by circles in Fig. \ref{map}. As one can see, all considered stellar clumps are the regions of current or very recent star formation. At the same time, BER1,2 overwhelm other young stellar complexes by luminosity and probably by stellar mass significantly.

What caused a simultaneous burst of star formation in two small regions observed at close branches of a bifurcated spiral arm, remains a puzzle. It is possible that they are not unique, we just catch two star-forming regions in the most luminous phase. These compact regions could not be the fragments of a dwarf galaxy crossing a disc of the parent galaxy, inspiring a strong shock wave and gas compression,  because BERs seem to be connected with spiral arms, and they do not reveal a significant velocity shift with respect to the adjacent regions.  
It is worth noting that in the northern half of the galaxy where BERs are observed, a spiral pattern is significantly smoothed or destroyed. As far as we don't see any companion which evidently may be responsible for the distortion of structure of NGC 5829,  the possibility remains that the latter is a result of recent  accretion of  gas falling onto a northern part of a disc from outside. The accreting gas could  strongly enhance star formation in the regions where the flow meets the dense native gas concentrations in spiral arms.  However, the oxygen abundance radial profile does not show any signs of dropping in the region of BERs, which in the case of accretion  could indicate that either a falling gas is   pre-enriched, or the  mass of absorbed gas is too small to affect chemistry. More detailed data on the neutral gas dynamics and distribution are needed to verify it. 

The spectra of BER1,2 stand out for their high [O~\textsc{iii}]/H$\beta$ ratios while having rather moderate [S~\textsc{ii}]/H$\alpha$ and [N~\textsc{ii}]/H$\alpha$ ratios. It seems that the \HII regions around both stellar superclusters containing massive stars are density bound, so that a significant fraction of ionizing photons is leaking in the surrounding gas. 

Curiously, all three emission regions crossed by the slit PA=30\degr, including BER1, reveal  velocity gradients  within the range 5--7~arcsec (2--3 kpc), which may be interpreted as the rotation with respect to the surrounding disc at the same direction as the disc as whole. It is compatible with the idea that these regions were formed by the compression of rotating gas in the disc. Still it is not evident that the gas is gravitationally bound at this scale. Note however that if to admit that these regions are close to be virialized, their masses should be an order of a few $10^8M_\odot$, which is compatible with the mass estimate of stellar population in BER1.  

The second spectral cut at PA=82\degr~ does not demonstrate any significant velocity gradient across the similar-looking emission region BER2. If both young massive stellar complexes -- BER1 and 2 -- have a  similar age and the formation mechanism, which seems natural, then their observed kinematic differences evidence  that they do not present a gravitationally coupled systems at kpc-scale. Of course, there remains a possibility that they are  strongly flattened and  dynamically isolated configurations observed at different inclination angles (nearly face-on in the case of BER2).

It is not surprising that the  massive stellar complex distorts the velocity field around it. Indeed, our observations demonstrate the enhanced non-circular gas motions within the region of several kpc-size around BER1, however similar behavior of LOS velocities also takes place in   another bright emission region crossed by slit PA 30\degr ~closer to the centre, in spite of the absence of compact stellar source there. 

The second star-forming clump, BER2, does not demonstrate any  velocity peculiarity.  So far as there is no ground to consider BERs as dynamically detached regions. They belong to a disc population and formed recently in the spiral arm branches as the result of some strongly non-stable process.

\subsection{Arp~82}

 Total luminosity $L_r$ of NGC~2535 is 1.6$\times 10^{10}L_\odot$ , and the stellar mass $M_*$   corresponding to its colour {\it (g-r)}, obtained from SDSS image, is  5.6$\times10^{10}M_\odot$   for \citet{Bell2003} model of stellar population. Similarly found stellar mass for NGC~2536 is 8.7$\times10^9M_\odot$. Based on the mass-metallicity relation described in \citealt{Sanchez2017} for the O3N2 calibrator, these mass values agree with the oxygen abundances of galaxies (8.5 - 8.6 for NGC~2535 and about 8.4 for NGC~2536). 

Our observations confirm the regular rotation of NGC 2535 and  a smooth velocity distribution along the bridge, connecting two galaxies. In contrast,  there is a strong non-circular gas motion in the satellite galaxy -- both in the central part and in the spiral arm adjacent to the bridge. At the same time the velocity dispersion of emission gas, estimated from the linewidths, remains low along the bridge, evidencing the laminar flow of gas, being a little shifted by velocity from the stellar component of the bridge.  

\begin{figure}
\hspace{-1.2
cm}
\includegraphics[width=1.3\linewidth]{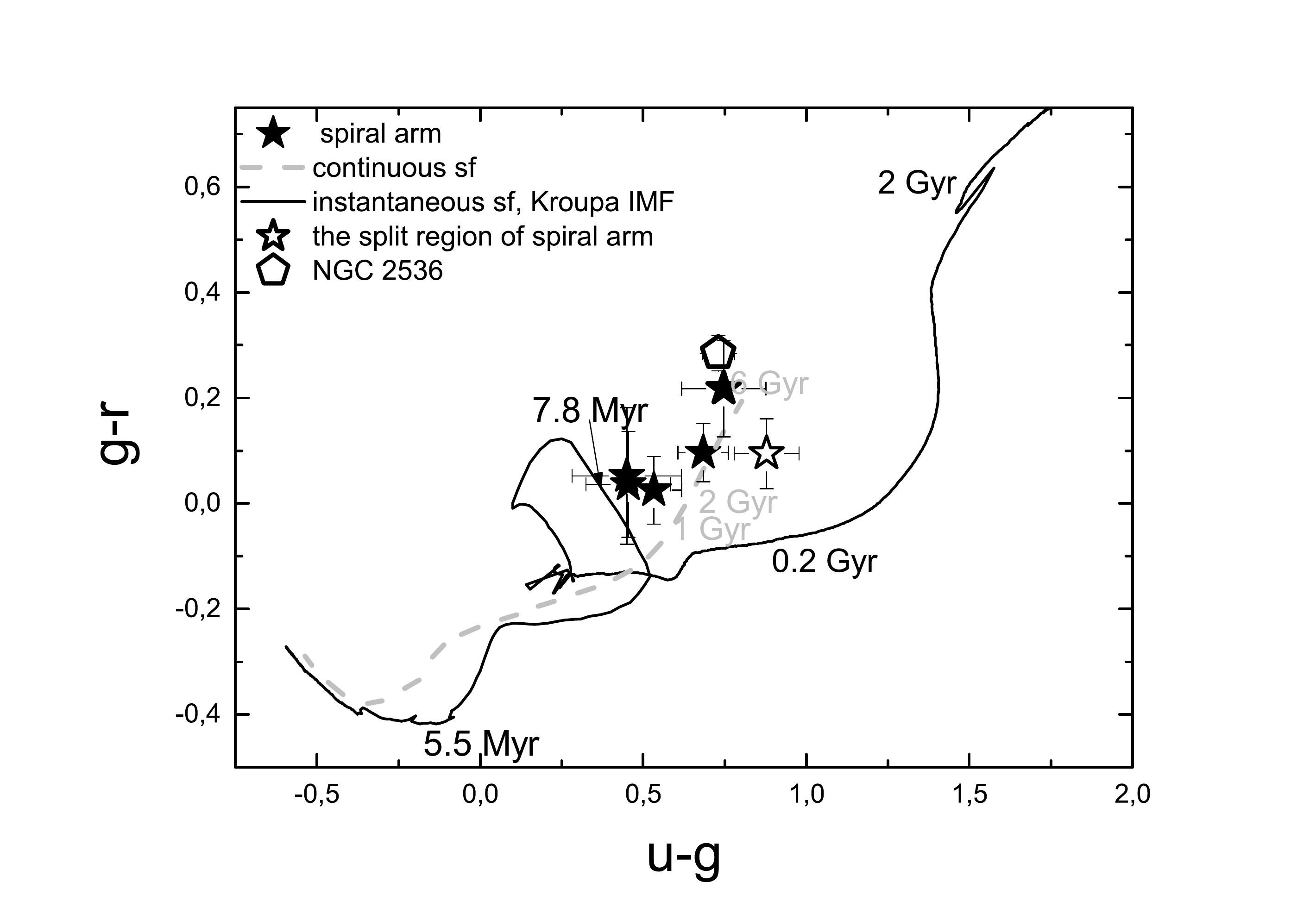}
\caption{The (g-r) vs (u-g) diagram. Filled stars show the position of star forming sites in the spiral arm of NGC~2535. Open star relates to the diffuse region of the spiral bridge. Open hexagon gives the position of the small companion galaxy NGC~2536. Black and gray lines demonstrate  Starburst99 model tracks for instantaneous and continuous star formation respectively.  We mark the ages of stellar population corresponding to the closest regions of the track by black and gray printing type for instantaneous and continuous star formation models correspondingly.}
\label{arp82_col}
\end{figure}

To find out the age of stellar population in the starforming sites of  Arp~82, in  Fig.\ref{arp82_col}  we show the position of selected emission regions crossed by the slits between the main body of NGC2535 and its satellite, as well as for the central part of NGC2536 (identified with circles in Fig. \ref{map}) at the (g-r) vs (u-g) diagram. Colours were estimated from the photometry of SDSS images and corrected for  extinction calculated from the $H_\alpha/ H_\beta$ ratio: $E_{B-V}\approx 0.15$. The local sites of star formation in the spiral arm (bridge) are shown by stars, the hexagone gives the position of the satellite. Open star corresponds to the diffuse region between galaxies where the bridge spiral arm seems to be splitting (see below). 

Evolution tracks in the Fig. \ref{arp82_col} are shown  for two models: instantaneous star formation (black line) and constant star formation rate 
(gray dashed line). Colours of the regions encircling blue stellar clumps and diffuse emission in the spiral arm (bridge)  marked by  stars are in good agreement with the model track for instantaneous star formation and correspond to the prevailing input of the light of young stars, so their luminosity-weighted age is less than 10 Myr (most probably, in the range 7.5-8 Myr). 

The situation is different for the other emission regions. One of them, denoted by open star, is situated in the low brightness bifurcated part of spiral arm  NGC2535. The other two regions lie in the bridge: one is the closest to NGC2535 and the second relates to the  diffuse emission region close to NGC~2536. These three areas are better described by the model with continuous star formation and  definitely contain  not only young, but also the older stars: their colours agree with  constant star formation rate during the last several billion yr (2-6 Gyr). The same is correct for the central  region of NGC 2536 which definitely has a large age.

As it follows from the abundance profile for PA=147\degr ~ there is a jump of (O/H) at about 0.2 dex between the bridge and the off-center regions of the main body of a small galaxy NGC~2536  (the abundance of gas in the satellite is higher than in the adjoining regions closer to the main galaxy).  It allows to propose that the gaseous bridge does not mix with the gas observed in the satellite. The same conclusion follows from the spectral cut at PA=342\degr: the oxygen abundance in the satellite is higher than in  the neighbour extended  emission "island" located closer to the main galaxy. This region belongs to the bridge rather than to the satellite, because its LOS velocity lays at the continuation of the velocity profile of NGC~2535, and differs at about 80 \kms\ from the central velocity of the satellite.  The oxygen abundance of this region  coincides with that observed in the bridge (see Fig.\ref{arp82_results}). 

The absence of clear signs of gas exchange between galaxies agrees with Klaric' dynamic model of the  interacting system,  which shows   that the bridge intersects NGC 2536 only in projection, so the satellite is a little closer to us \citep[see][]{Kaufman1997}. In this case the enhanced star formation in the satellite is not a result of collision with the gas flowing along the bridge,  but rather is caused by internal processes triggered  by tidal disc disturbance.

\begin{figure}
\includegraphics[width=\linewidth]{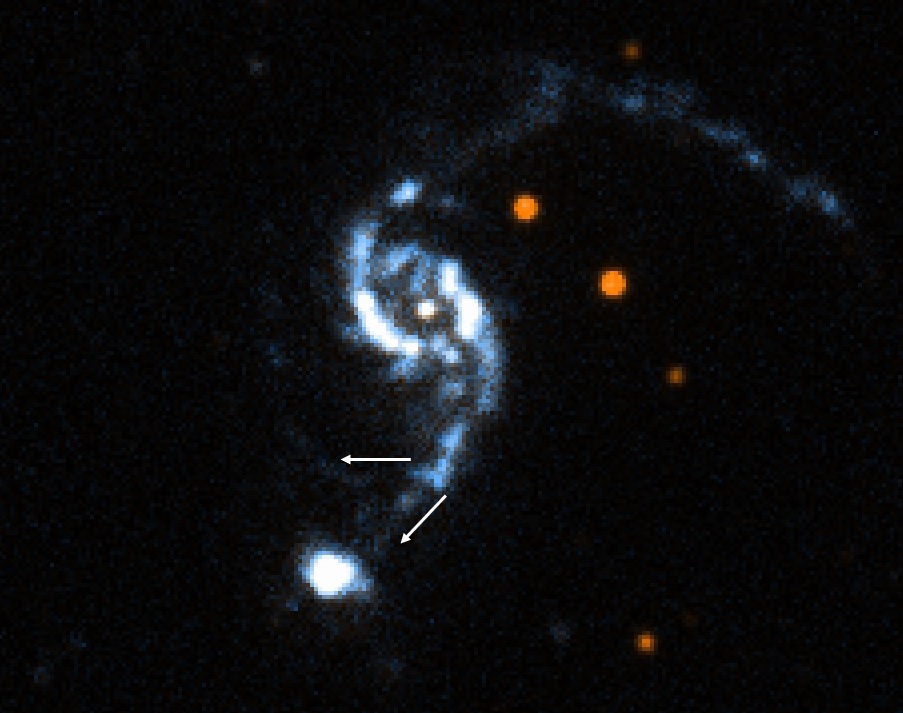}

\includegraphics[width=\linewidth]{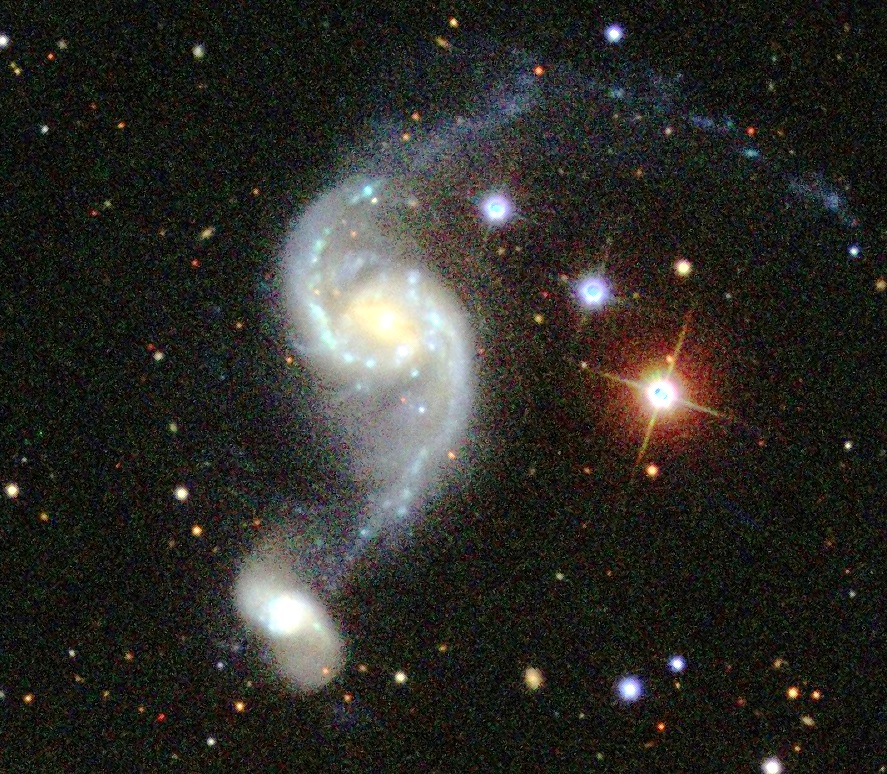}
\caption{Combined GALEX FUV and NUV-bands with overplotted arrows showing the position of bifurcated spiral arms (top) and SDSS images of Arp 82 (bottom).}
\label{arp82_uv}
\end{figure}

The other interesting feature of NGC~2535 is the faint trace of the third spiral arm, which branches off the  bright arm running to the satellite. First it was noted by \citet{Hancock2007}. A faint branch is easily visible at the SDSS  and GALEX images in Fig.\ref{arp82_uv}, where the direction of branching spiral arms are marked by arrows. This narrow arm possesses blue color and looks symmetric with respect to opposite, northern arm.  It seems that we see the residual arm of  symmetric (in the past) spiral structure of the galaxy, existed before the encounter happened. Most probably,  a tidal force imposed a new mode of the density wave in the disc, which gave rise to the arm-bridge, while the previous  mode had not yet time to die totally out.  Note in advance that similar feature is also  noticeable in Arp~58 (see the next subsection).
\subsection{Arp~58}\label{arp58_discussion}
\begin{figure}
\includegraphics[width=\linewidth]{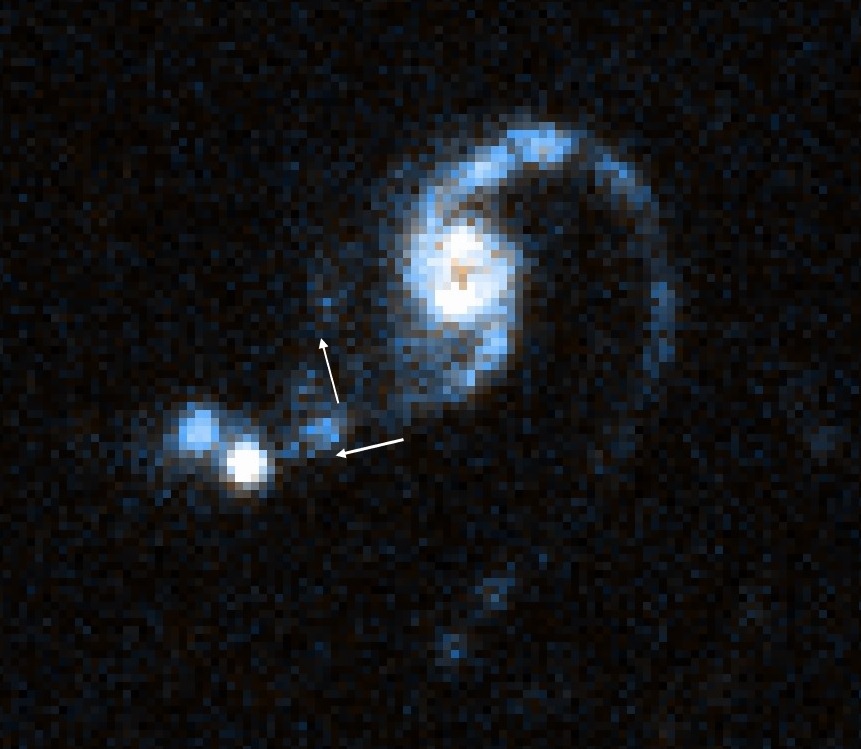}
\caption{Combined GALEX FUV and NUV-bands image of Arp 58 with overplotted arrows showing the position of bifurcated spiral arms.}
\label{arp58_uv}
\end{figure}

In many respects this system is similar to Arp~82, described in the previous section, although it is less studied,  and the masses ratio of the main galaxy and the satellite in Arp~58 is significantly higher than in the former case. In both systems we have a long tidal tail of the main galaxy and a bluish bridge-like spiral arm, connecting two galaxies. Most probably, this arm  does not end on  the satellite galaxy, but rather pass a little further projecting onto it.  For Arp~82 a similar proposal was argued above, and in the case of Arp~58 such possibility is based on  the colour image of the system, especially in UV light (see the combined FUV and NUV GALEX image in Fig.\ref{arp58_uv}), where the blue spiral arm looks projecting on the satellite galaxy. In this case a  blue ejection-like feature emanating from the body of a satellite may be considered as the continuation of UV bridge-arm of the main galaxy. A noticeable  dark gap between this feature and the satellite is especially prominent in the UV GALEX image, which, if it is accounted for the dust extinction in the satellite, evidences that the bridge arm is behind  the satellite.

Another curious detail that is evident from the UV-image is the faint offshoot of the spiral arm-bridge  at about a half-way between the centers of the galaxies.  This thin arc-like spur  looks  symmetric with respect to the opposite arm of the main galaxy, similarly to that in Arp~82.  

\begin{figure}
\includegraphics[width=2.3\linewidth]{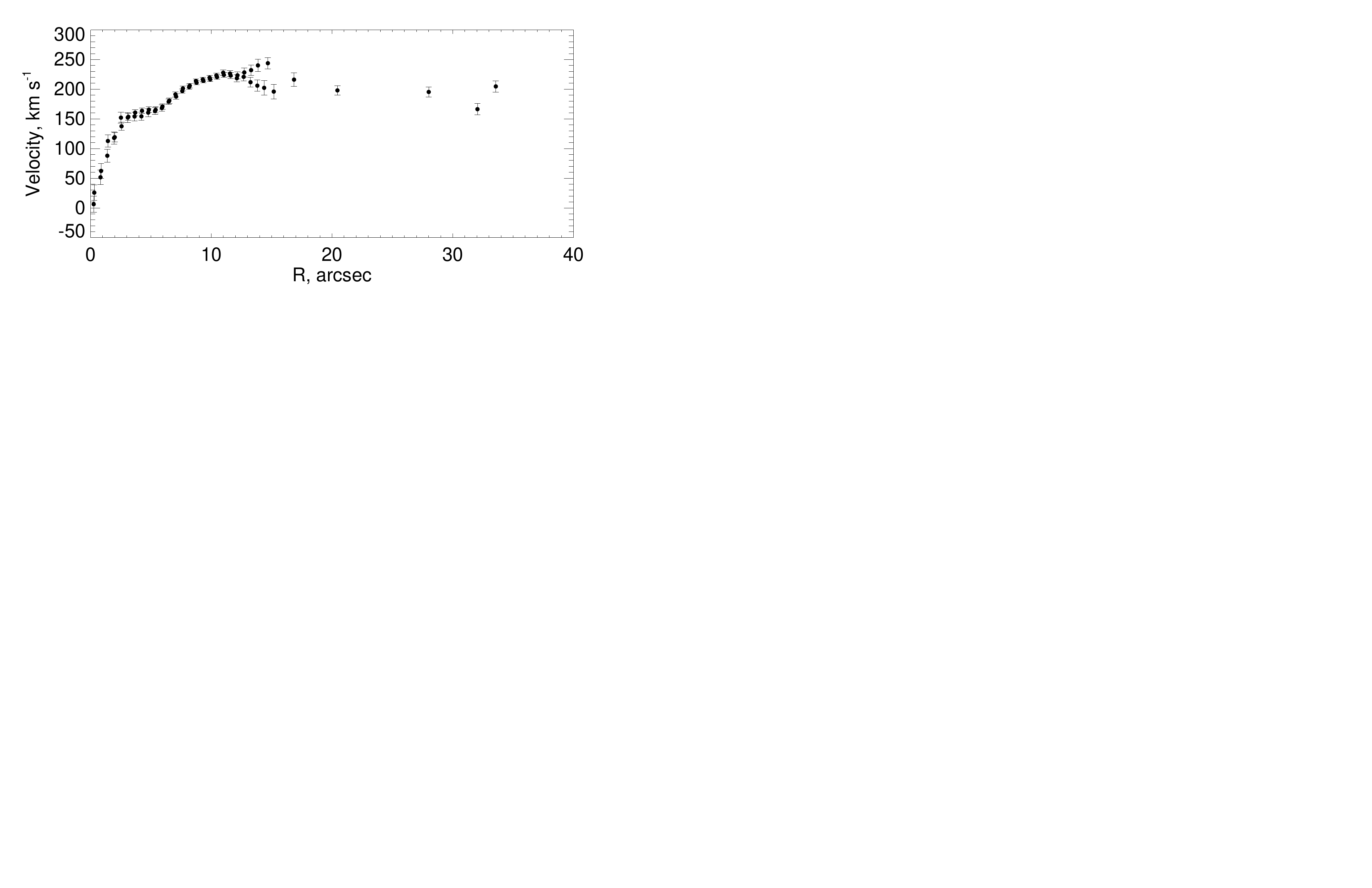}
\vspace{-9.0cm}
\caption{The $H_\alpha$ rotation curve of Arp~58.}
\label{arp58_rc}
\end{figure}

 A symmetric gas velocity profile of UGC~4457 allows to restore its rotation velocity curve. It is presented in Fig \ref{arp58_rc} for the adopted  inclination angle $i = 65$\degr  ~and position angle $(PA)_0=164.4$\degr ~taken from HYPERLEDA. The curve grows slowly, reaching maximum or flat region of the curve at  about 11--12~arcsec,  although the presence of non-circular gas motion is  evident at larger distances. The `knee' (a change of gradient of gaseous velocity curve) at  R$\sim$ 3--6~arcsec most probably does not reflect a mass distribution: at these distances the slit crosses the bright spiral arm where radial non-circular velocities are quite expected.

 The main galaxy in Arp 58 does not possess a noticeable bulge, so the upper limit of mass of stellar population of the disc may be estimated if to admit that its circular velocity does not exceed the maximal measured velocity of rotation, reaching $V_{max}\sim$ 230 \kms at R=R$_{max}$ $\approx$ 12~arcsec. For a disc with exponential mass distribution it corresponds to the dynamic mass $M_{disc}$<$ 1.2V_{max}^2R_{max}/G$ =  1.26$\times10^{11}M_\odot$. Accounting for dark halo may reduce this value. This limit agrees with the mass of stellar population based on the photometry of SDSS image which gives a slightly lower mass value. Indeed, we found the total luminosity of galaxy $L_r\approx 3.2\times10^{10}L_\odot$ and the observed $g-r = 0.68$ (corrected for Galactic extinction), corresponding to $M_*/L_r\sim$2.8 solar units according to the  model of \citet{Bell2003},  which gives  $M_*\approx 9\times10^{10}M_\odot$. Here the internal extinction is ignored; taking it into account will reduce the mass estimate due to the lower colour index and $M_*/L_r$ ratio, however it will be  compensated by the decreasing of $L_r$. Note that the estimation of dynamical  mass is also very preliminary.  The additional velocity measurements are needed to clarify $(PA)_0$ of major axis and the inclination angle of the disc.
 
The emission regions between two galaxies, crossed by the slit, keep their LOS velocity close to those observed in the adjacent part of UGC~4457. Unlike UGC~4457, the satellite does not reveal a regular rotation: its half faced to the main galaxy demonstrate a zero LOS velocity gradient which  is evidently the result of gravitational interaction. It does not allow to get the dynamic mass. Photometric method which uses SDSS images ({\it g} and {\it r}-bands) gives for a small galaxy $M_*\approx 8.5\times10^8M_\odot$ and $M/L_r$ = 0.71 solar units (not corrected for inner absorption). 

Line intensity ratios at the BPT diagrams show that the emission regions in Arp 58 correspond to H~\textsc{ii}-spectra of photoionized gas and -- in some regions -- to the composite spectra. A special case is the single outlier at the diagrams in Fig.\ref{arp58_bpt} for Arp 58, revealing the highest ratios of ion-to-H$\alpha$ lines intensities. It belongs to the extended emission region in the center of the satellite galaxy, and may be considered as the indirect evidence of the moderate AGN-activity.
\subsection{Oxygen abundance gradients}\label{oh}
\begin{figure}
\includegraphics[width=\linewidth]{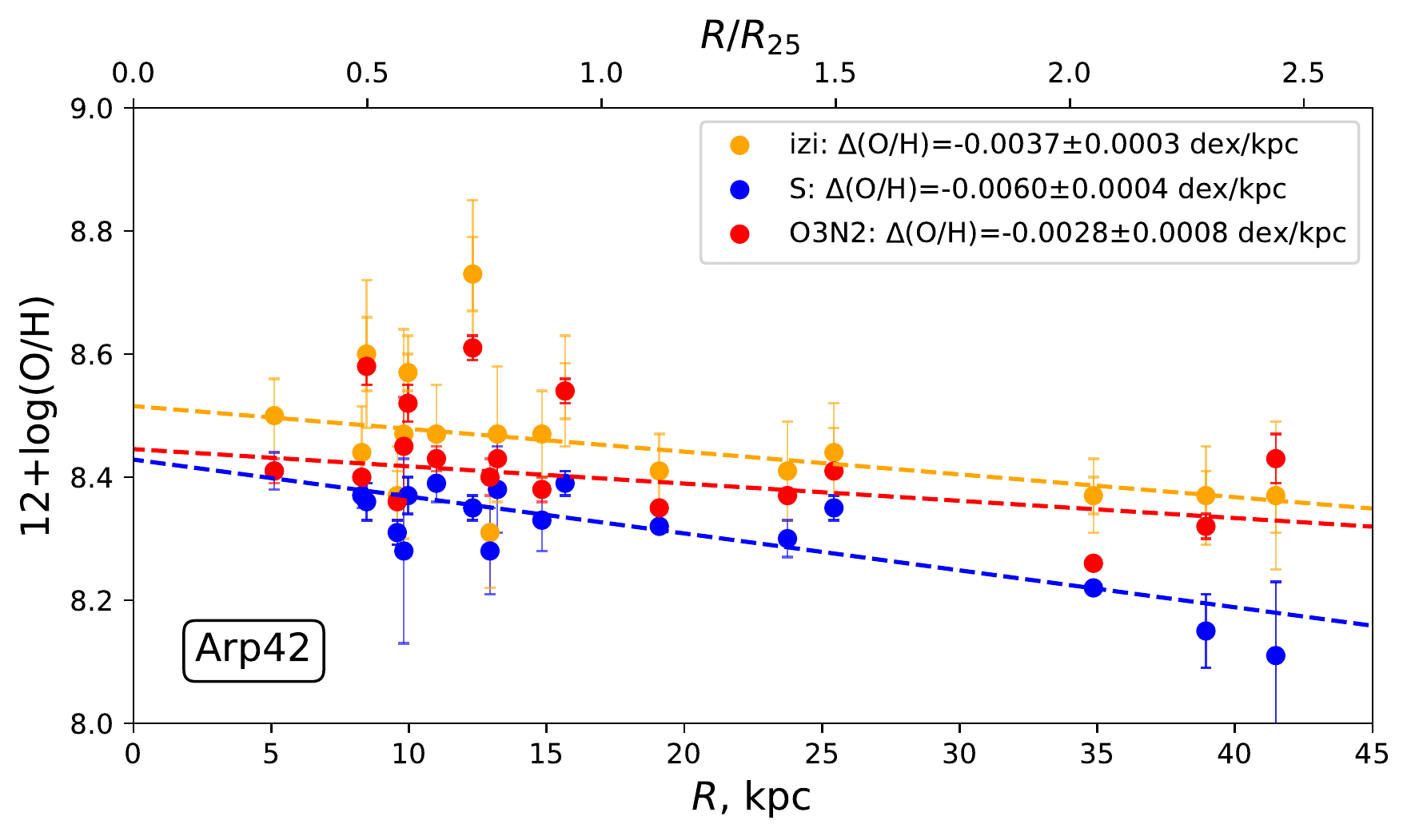}

\includegraphics[width=\linewidth]{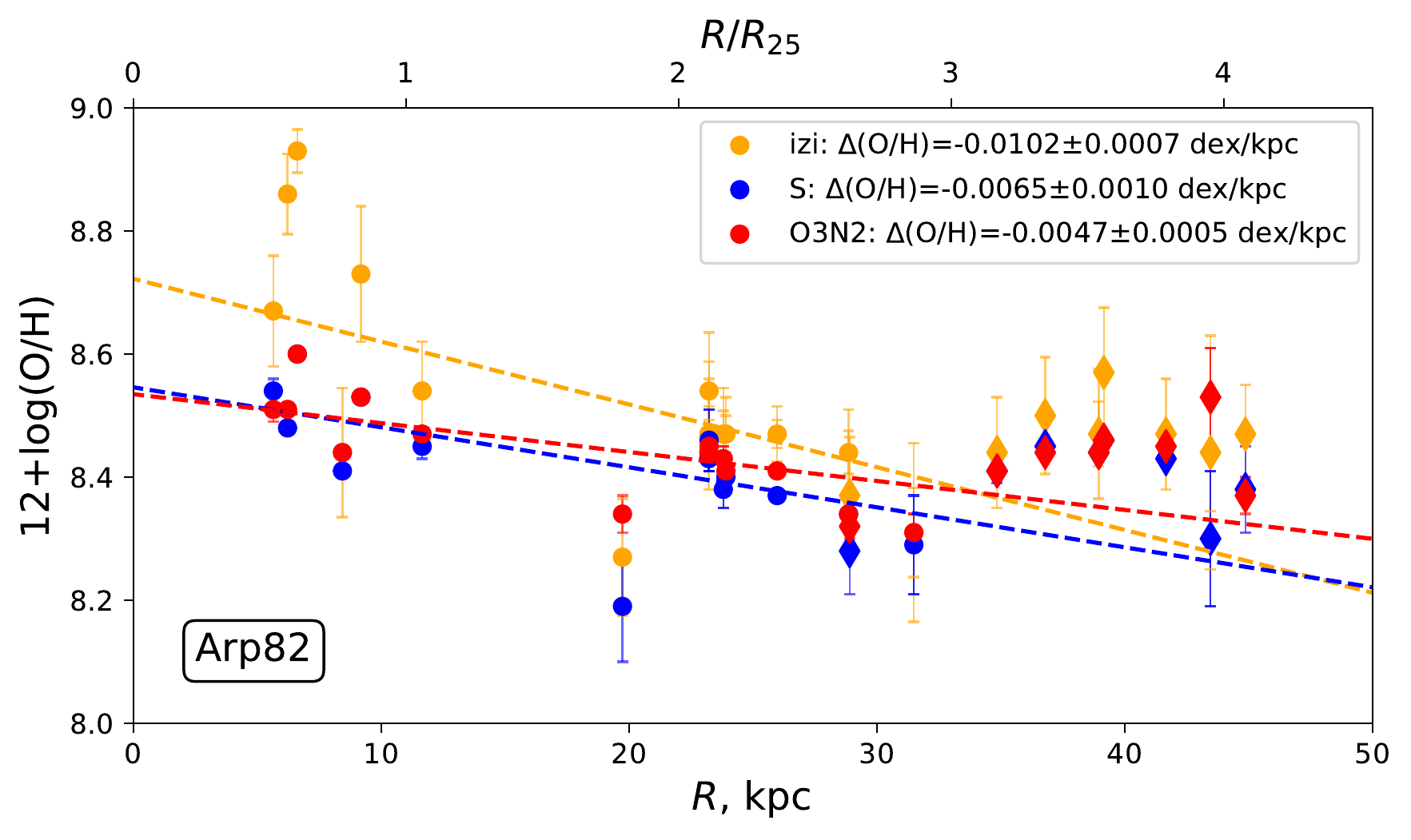}

\includegraphics[width=\linewidth]{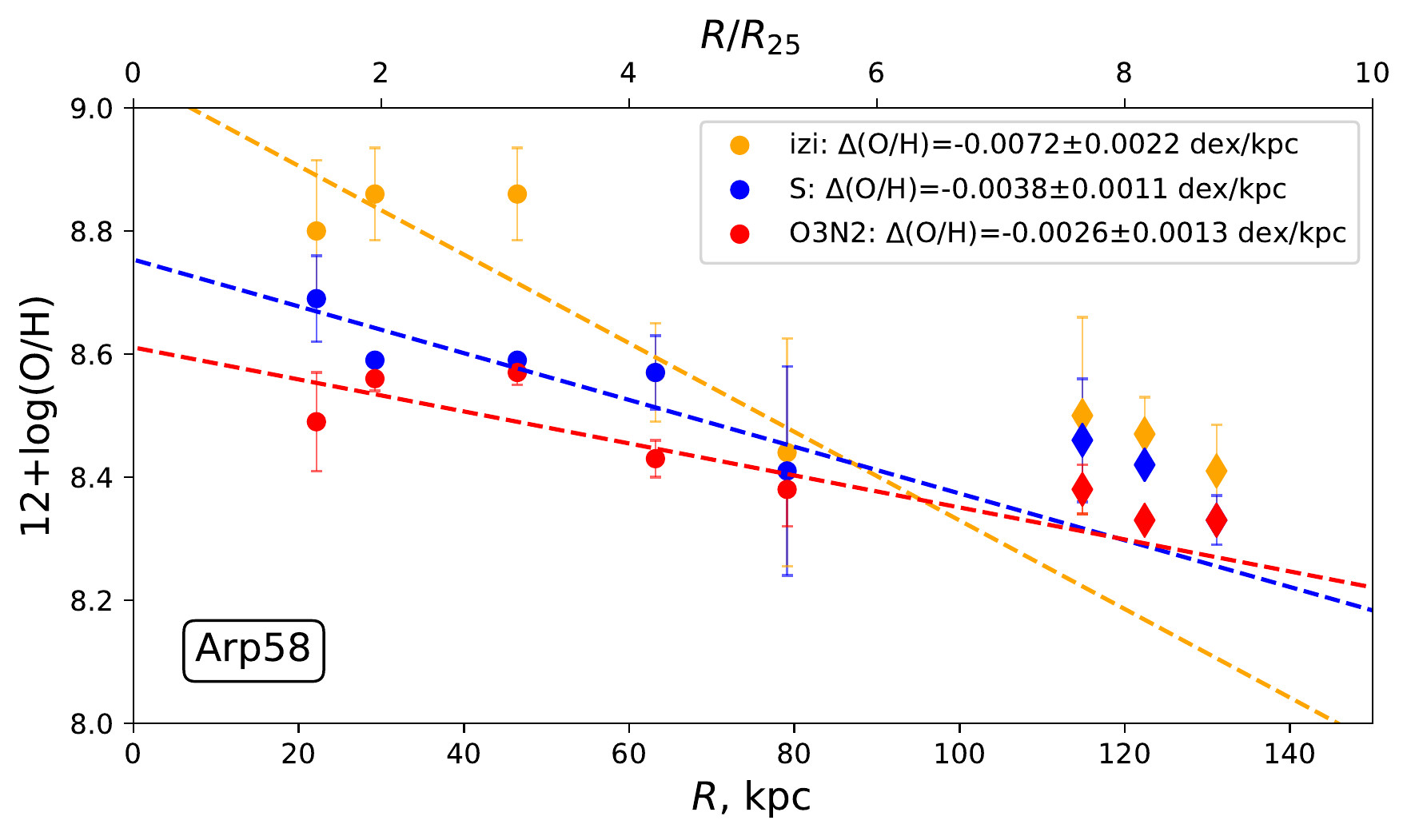}
\caption{From top to bottom: the radial profiles  of oxygen abundance of Arp~42, Arp~82 and Arp~58 (reduced to the disc plane of the main galaxy). The gradients were estimated by robust least-square fitting of the values for a disc of the main galaxy. The points related to the satellites are denoted by diamonds (assuming that they lay in the disc plane).}
\label{abund}
\end{figure}

   Fig. \ref{abund} shows the profiles of oxygen abundance as a function of the inclination-corrected  distance from the centre of the main galaxy  (reduced to its disc plane for the adopted inclination angle) and their linear fits  for all three galaxies. This figure excludes all  points with the suspected significant contribution of non-photoionization mechanism of excitation according to their positions at the diagnostic diagrams. Here we remind the reader that all regions with the suspected DIG contribution having EW(H$\alpha$) < 3 have also been  excluded from the analysis. In order to rely on the high S/N data we use the estimates  obtained for stacked spectra only (squares in Figs.~\ref{arp42_results}, \ref{arp58_results}, \ref{arp82_results}) for estimating the metallicity gradients. 

In all three cased the gradients are shallow, not exceeding -0.01 dex kpc$^{-1}$. It agrees with the lower abundance gradients
usually found for galaxies in close 
pairs \citep{Kewley2010,  Rosa2014}  and for interacting or merging galaxies \citep{Rupke2010,Rich2012} in comparison with single
galaxies, which may be accounted for the radial mixing of
gas inspired by interaction. It is interesting, that at large radial distances the flattening of gradient is often observed even for non-interacting galaxies \citep{Sanchez2014,califa2016, Sanchez-Menguiano2018}, that is  the mean abundance gradient decreases when the outskirts of galaxies are taken into consideration.  Two of the studied  systems reveal similar behaviour. The abundance gradients we estimated were measured along the radial coordinate (corrected for the disc inclination), including the regions of  tidal bridges in Arp 82 and Arp 58. Taken
separately, these outer regions do not reveal a clear evidence of any gradient of O/H, which explains the flattening of the abundance profiles.  However, we need more observations at this radial range to confirm it.

 It is known that a spread of (O/H) gradients in galaxies are lower when normalized to the physical radial scale such as the isophote radius $R_{25}$ or effective
radius $R_e$ \citep{Sanchez2014,califa2016, Sanchez-Menguiano2018}. Indeed, the mean gradients for three systems we consider become very similar if to express them in terms of $R_{25}$: they lay between $-0.05$ and $-0.1$ dex $R^{-1}_{25}$, which is significantly lower than for normal spiral galaxies ($-0.3-0.4$ dex $R^{-1}_{25}$ according to \citealt{Pilyugin2014, Ho2015}).

\section{General conclusions}

Our long-slit observations confirm that the systems we studied are dynamically disturbed and actively star-forming galaxies. They contain many emission knots (clumps) together with the extended emission regions of low brightness. The mean ages of stellar clumps found from {\it ugr}-photometry are very low and (with a few exceptions) exclude a notable contribution of old stellar population in their luminosity.  
A special interest present two extremely luminous compact clumps (BER-1,2) in the bifurcated spiral arm in Arp 42, which origin remains a puzzle. Their luminous masses within the scale of 1-2 kpc should  exceed $10^8~M_\odot$  unless their stellar IMF is not "bottom-light". The interaction with some  `dark' dwarf galaxy or the gas  accretion triggering the formation of super stellar clusters may not be excluded. 

LOS velocity profiles in the viewed galaxies reveal local regions of non-circular gas motions, which usually coincide with  bright star-forming sites. M51-type systems Arp 82 and Arp 58 demonstrate a regular rotation of the main galaxies, a low velocity gradient along bridges and a strong non-circular motion of gas in their satellites evidently caused by interaction. In both systems there observed faint arc-like offshoot from the bridge spiral arms, looking symmetrical with respect to the opposite bright arms of the main galaxy.  It allows to propose that in these galaxies we observe a transient two-mode wave spiral structure:  while a spiral arm looking as a bridge between galaxies was induced by tidal interaction, the   original spiral arm which existed before the close encounter, faded, so that we can observe only its weak trace.

At the diagnostic diagrams the low brightness areas are situated on both sides of the demarcation line, separating the area of stellar photoionization from shock excitation or hard radiation area. However in all systems we consider the low brightness emission regions prevail among those demonstrating the composite spectra, especially standing out for their enhanced [N~\textsc{ii}]/$H\alpha$ ratios. It reveals the presence of low brightness diffuse ionized gas in the systems, for which a harder ionization is needed to account for the line ratios. (Note that these regions were not used for the abundance estimations).  

Among the commonly invoked sources of additional ionization are the radiation of old hot pAGB stars, the leakage of hard photons from the star-forming regions, 
and the shock waves \citep[see discussion in][]{Zhang2017}. The first mechanism is
less likely because of very low brightness of old stellar population in the considered regions. The leakage of hard photons from the unresolved \HII regions looks quite acceptable.  The most outstanding case of LINER-like spectrum  is the nucleus of the satellite galaxy in Arp 58, where the line ratios give evidence of  the low level of AGN activity in this small galaxy, inspired by gravitational perturbation.

The presence of shock waves is natural to expect in the interacting systems. Indeed, shock waves  caused by interaction of tidally expelled gas with the ambient circumgalactic gas,  or by  collisions of disturbed gas flows in tidal debris,  may  trigger star formation beyond the main bodies of galaxies even in the regions of initially low gas density. It agrees with numerical simulations, which confirm that the strong compression and shocks produced by the
galaxy interaction fuel rapid formation of  stellar clusters after the first close pericentric passage \citep[see e.g.][]{Maji2017}.

The limited spectral and spatial resolution of our observations did not allow to reveal shock waves directly. A non-direct evidence  of their presence may be the enhanced velocity of gas turbulent motions in the regions, where the velocity dispersion exceeds the value 20--40 \kms, usually observed in \HII regions of galaxies. The existing model estimates show that the enhanced  velocity dispersion of emission gas may account for the line ratios of the composite LINER-like spectra at the diagnostic diagrams \citep[see e.g.][]{Rich2011} in agreement with observations of galaxies \citep{Oparin2018}.  

However in the case of our systems the regions where the velocity dispersion reaches or exceeds 50 \kms\, are rare.  Most evidently, they include  the central regions of both galaxies and the bright area in the base of spiral bridge between galaxies in the system Arp 82. As expected, their positions at BPT diagram reveal a composite spectra. 

All considered galaxies  have  low gradient of (O/H) along the radius of central galaxy which does not exceed $-(0.003-0.005)$ dex~kpc$^{-1}$ for Arp~42 and Arp~58 and $-(0.005-0.01)$ dex~kpc$^{-1}$ for Arp~82 if not to count its satellite	(see Fig. \ref{abund}). If to express the gradients in terms of $R_{25}$: they lay between $-0.05$ and $-0.1$ dex $R^{-1}_{25}$, which is significantly lower than for normal spiral galaxies ($-0.3-0.4$ dex $R^{-1}_{25}$ according to \citealt{Pilyugin2014, Ho2015}). It allows to conclude that the gas in their discs was partially expelled outwards as the result of interaction (at least for the last two galaxies). However the oxygen abundance of the central regions of galaxies we consider in general agrees with the "metallicity-stellar mass" dependence for non-interacting galaxies.  There is no clear evidence of gas collision or gas chemical mixing between the main galaxies and their satellites in M~51-type systems Arp~82 and Arp~58.

\section*{Acknowledgements} 
The authors are grateful to anonymous referee for valuable comments that helped to improve the manuscript. 
They also thank V.Afanasiev and R.Uklein for a collaboration. The observations at the 6-meter BTA telescope were carried
out with the financial support of the Ministry of Education
and Science of the Russian Federation (agreement No. 14.619.21.0004, project ID RFMEFI61914X0004).
 The reduction of the spectral data were supported by The Russian Science Foundation
(RSCF) grant No. 17-72-20119. The estimate of the metallicity and its interpretaion were supported by Russian Foundation for Basic Research grant No. 18-32-20120. The authors acknowledge the usage of the
HyperLeda database (http://leda.univ-lyon1.fr). Authors acknowledge support from the Leading Scientific School in astrophysics (direction: extragalactic astronomy) at Moscow State University.
Funding for the Sloan Digital Sky Survey IV has been provided by the Alfred P. Sloan Foundation, the U.S. Department of Energy Office of Science, and the Participating Institutions. SDSS-IV acknowledges
support and resources from the Center for High-Performance Computing at
the University of Utah. The SDSS web site is www.sdss.org.

SDSS-IV is managed by the Astrophysical Research Consortium for the 
Participating Institutions of the SDSS Collaboration including the 
Brazilian Participation Group, the Carnegie Institution for Science, 
Carnegie Mellon University, the Chilean Participation Group, the French Participation Group, Harvard-Smithsonian Center for Astrophysics, 
Instituto de Astrof\'isica de Canarias, The Johns Hopkins University, 
Kavli Institute for the Physics and Mathematics of the Universe (IPMU) / 
University of Tokyo, Lawrence Berkeley National Laboratory, 
Leibniz Institut f\"ur Astrophysik Potsdam (AIP),  
Max-Planck-Institut f\"ur Astronomie (MPIA Heidelberg), 
Max-Planck-Institut f\"ur Astrophysik (MPA Garching), 
Max-Planck-Institut f\"ur Extraterrestrische Physik (MPE), 
National Astronomical Observatories of China, New Mexico State University, 
New York University, University of Notre Dame, 
Observat\'ario Nacional / MCTI, The Ohio State University, 
Pennsylvania State University, Shanghai Astronomical Observatory, 
United Kingdom Participation Group,
Universidad Nacional Aut\'onoma de M\'exico, University of Arizona, 
University of Colorado Boulder, University of Oxford, University of Portsmouth, 
University of Utah, University of Virginia, University of Washington, University of Wisconsin, 
Vanderbilt University, and Yale University.

\bibliographystyle{mnras}
\bibliography{arp42}
\label{lastpage}
\end{document}